\documentclass{jfm}

\usepackage{mathtools}                    %
\usepackage{mathrsfs}                   %
\usepackage{stmaryrd}                   %
\usepackage[symbol,perpage]{footmisc}
\usepackage[numbers]{natbib}
\usepackage[np,autolanguage]{numprint}
\usepackage{color}
\usepackage{afterpage}
\usepackage{setspace}
\usepackage{enumerate}
\usepackage{balance} 
\usepackage{calc}
\usepackage{caption}
\usepackage{subcaption}
\usepackage{textcomp}
\usepackage{bm}
\usepackage{graphicx}
\usepackage[latin1]{inputenc}
\usepackage{xspace}
\usepackage{rotating}
\usepackage{hyperref}
\usepackage{upgreek}
\usepackage[colorinlistoftodos]{todonotes}

\usepackage{mathtools}
\usepackage[np,autolanguage]{numprint}
\usepackage[normalem]{ulem}
\usepackage{acronym}
\acrodef{ALE}[ALE]{Arbitrary Lagrangian Eulerian}
\acrodef{FSI}[FSI]{fluid-structure interaction}
\acrodef{FEM}[FEM]{finite element method}
\acrodef{BVP}[BVP]{boundary value problem}
\acrodef{BVPs}[BVPs]{boundary value problems}
\acrodef{GLM}[GLM]{Generalized Lagrangian Mean}
\acrodef{DAE}[DAE]{differential-algebraic equations}
\acrodef{BDF}[BDF]{backward differentiation formulas}
\acrodef{BAW}[BAW]{bulk acoustic wave}
\acrodef{SAW}[SAW]{surface acoustic wave}
\acrodef{PDMS}[PDMS]{polydimethylsiloxane}

\DeclareMathOperator{\Lin}{\ensuremath{\mathscr{L}(\mathscr{V},\mathscr{V})}}
\DeclareMathOperator{\trace}{tr}

\newcommand{\comsol}{COMSOL Multiphysics\textsuperscript{\textregistered}\xspace}

\renewcommand{\d}[1]{\ensuremath{\mathrm{d}#1}}
\newcommand{\bv}[1]{\vec{#1}}
\newcommand{\tensor}[1]{\mathsf{#1}}    %

\newcommand{\transpose}[1]{#1^{\scriptscriptstyle\mathrm{T}}}
\newcommand{\inversetranspose}[1]{\ensuremath{#1^{\scriptscriptstyle-\mathrm{T}}}}
\newcommand{\scdot}{\ensuremath{\!\cdot\!}}

\newcommand{\microns}{\ensuremath{\,\upmu\text{m}}}
\newcommand{\order}[3][]{\ensuremath{#2^{{#1}\scriptscriptstyle(#3)}}}
\newcommand{\orderT}[3][]{\ensuremath{#2^{{{#1}\scriptscriptstyle(#3)}^{\scriptscriptstyle\mathrm{T}}}}}

\newcommand{\PFF}{\ensuremath{\tensor{F}^{\scriptscriptstyle(0)\#}}}
\newcommand{\PFu}{\ensuremath{\bv{u}^{\scriptscriptstyle(0)\#}}}
\newcommand{\colondot}{\mathbin{:}}
\newcommand{\LM}[1]{\ensuremath{#1^{\mathrm{L}}}}
\newcommand{\PF}[1]{\ensuremath{#1^{\scriptscriptstyle\#}}}

\newcommand{\EMC}{\ensuremath{\mathrm{E}_{\mathrm{mbc}}}}
\newcommand{\EZBC}{\ensuremath{\mathrm{E}_{\mathrm{zbc}}}}

\renewcommand{\vec}[1]{\bm{#1}}

\newcommand{\ii}{\mathrm{i}}

\newcommand{\beq}[1]{\begin{equation} \eqlab{#1}}
\newcommand{\eeq}{\end{equation}}
\newcommand{\bsub}{\begin{subequations}}	
\newcommand{\esub}{\end{subequations}}
\def\bal#1\eal{\begin{align}#1\end{align}}
\def\bsubal#1\esubal{\bsub \begin{align}#1\end{align} \esub}

\begin{document}

\title{Acoustic Streaming: An Arbitrary Lagrangian-Eulerian Perspective}

\author
 {
 Nitesh Nama\aff{1},
  Tony Jun Huang\aff{2}
  \and 
  Francesco Costanzo\aff{1, 3}
  \corresp{\email{costanzo@engr.psu.edu}}
  }

\affiliation
{
\aff{1}
Department of Engineering Science and Mechanics, The Pennsylvania State University, University Park, PA 16802, USA
\aff{2}
Department of Mechanical Engineering and Materials Science, Duke University, Durham, NC 27708, USA
\aff{3}
Center for Neural Engineering, The Pennsylvania State University, University Park, PA 16802, USA}

%
%
%
%

\maketitle

\begin{abstract}
We analyze acoustic streaming flows using an \ac{ALE} perspective. The formulation stems from an \emph{explicit} separation of time-scales resulting in two subproblems: a first-order problem, formulated in terms of the fluid displacement at the fast scale, and a second-order problem formulated in terms of the Lagrangian flow velocity at the slow time scale. Following a rigorous time-averaging procedure, the second-order problem is shown to be \emph{intrinsically} steady, and with \emph{exact} boundary conditions at the oscillating walls.  Also, as the second-order problem is solved directly for the Lagrangian velocity, the formulation does not need to employ the notion of Stokes drift, or any associated post-processing, thus facilitating a direct comparison with experiments. Because the first-order problem is formulated in terms of the displacement field, our formulation is directly applicable to more complex fluid-structure interaction problems in microacosutofluidic devices.
After the formulation's exposition, we present numerical results that illustrate the advantages of the formulation with respect to current approaches.
\end{abstract}

\section{Introduction}
\label{section: Introduction}
Over the past few decades, microfluidics has received significant interest owing to its numerous applications in biology, chemistry, and medicine~\citep{squires2005microfluidics,lee2013third}. In these fields, microfluidic devices and techniques offer significant advantages over the traditional ones in terms of miniaturization, low-cost, automation, precise microenvironment control, and reduced sample consumption~\citep{Ding2013}. However, there are significant challenges to overcome in the use of microfluidic devices in key applications that require pumping and mixing~\citep{squires2005microfluidics}.
For example, the flow at microscales is usually laminar, thereby precluding turbulent mixing, and lending itself only to slow diffusion-dominated mixing. To overcome these challenges, as well as to achieve novel functionalities, researchers have integrated different physics into the microfluidic platforms such as electrokinetics, chemistry, optics, biotechnology, etc. Among these, microacoustofluidics, which refers to the merger of acoustics and microfluidics, has shown great promise~\citep{Ding2013,friend2011microscale}. These systems are typically characterized by the propagation of high-frequency acoustic waves through fluids. The wave propagation not only induces a corresponding high-frequency response of the fluid, but also a slow mean flow through a nonlinear phenomenon referred to as \emph{acoustic streaming}.

Recently, several studies concerning the physical mechanism and mathematical modeling of acoustic streaming in microacoustofluidic systems have been reported \citep{koster2007numerical,muller2012numerical,nama2015numerical,Vanneste2011Streaming-by-Le0}. These studies typically employ Nyborg's perturbation approach wherein the Eulerian flow variables are split into their first-order and second-order components with respect to a suitably identified smallness parameter, typically based on a choice of length scale~\citep{Nyborg1998Acoustic-Stream0}. This results in a set of linear equations such that the first-order approximation of the balance laws determine the acoustic response of the systems, while the second-order equations, upon time-averaging, govern the time-averaged mean response of the fluid. To the authors' best knowledge, the vast majority of models based on this approach is developed within a fully Eulerian framework.  While conceptually acceptable, one difficulty with this framework is the fact that the boundary conditions are approximated by necessity.  The reason is that, aside from the physical justification underlying a specific boundary condition, microacoustofluidic devices are actuated via a high frequency excitation of the solid component of the system.  As such, a fundamental aspect of the boundary conditions is the \emph{displacement field} of the device's walls.  In a fully Eulerian context, the relationship between the fluid velocity and the wall displacement fields must be approximated, as demonstrated, for example, in \citet{koster2007numerical,muller2012numerical,Bradley1996Acoustic-Stream0,bradley2012acoustic}.

Another very important element that makes a fully Eulerian framework not entirely ideal pertains the experimental validation of models. Many experimental techniques in microfluidics rely on particle tracking.  As such, a comparison between theory and experiments requires the determination of the particle (or Lagrangian) velocity field, which, in time-periodic fluid flows and within an Eulerian framework, requires the notion of Stokes drift \citep{Buhler2009Waves-and-Mean-0}.  Consequently, the comparison of numerical predictions to experiments usually requires an additional layer of approximation associated with the post-processing of the Eulerian velocity field to obtain the Lagrangian mean-field flow. Although not intrinsically Eulerian-framework related, there is an additional aspect of the governing equations that are derived in this framework that deserves more careful consideration.  Specifically, the governing equations for the streaming flow are obtained via a time-average operation in which the exact relationship between slow and fast time scales is seldom made explicit.  This results in a slow-time-dependent system of equations for which steady solutions are typically sought.  Again, while not intrinsically problematic, implicit assumptions are hidden in the time averaging process and in its steady approximation that might impact the validity of current predictions and their difficulty in capturing key features of streaming flows.

Going back to the examination of the boundary conditions, we have remarked that the excitation of microacoustofluidic devices results from \ac{FSI}.  Hence, we decided to revisit the very formulation of the governing equations, this time using an approach frequently adopted in the formulation of \ac{FSI} problems. Furthermore, in revisiting the consequences of the time averaging operation, we noted a recent multi-scale analysis by~\citet{Xie2014multiscale} of the dynamics of a spherical particle in an acoustic field. This study is based on a time-scale separation technique that is very rigorous and makes time averaging more transparent, and in which, for certain flow regimes the inertial terms do not appear in the second-order equations, which is in contrast to the aforementioned Eulerian approaches.

In this work, we frame the acoustic streaming problem in an \ac{ALE} context where the first-order problem is formulated in terms of the fluid displacement. Following a multiscale approach, similar to that employed by~\citet{Xie2014multiscale}, the second-order system is then derived directly in terms of the Lagrangian flow velocity. We believe that the \ac{ALE} formulation, with no significant computational overhead compared to the commonly employed Eulerian formulation, offers several distinct advantages over the Eulerian formulation:  Firstly, the ALE formulation is consistent with the usually employed Eulerian formulation for a general time-dependent case. However, the time-averaging process in this formulation is much more transparent, leading naturally to a time-independent flow at the second-order. Secondly, since the first-order problem is formulated in terms of the displacement, this approach offers a natural extension to fluid-structure interaction problems concerning the motion of particles inside a microacoustofluidic device. Thirdly, the formulation of second-order problem in terms of Lagrangian flow velocity circumvents the need to utilize the concept of drift motion and the solution can be directly compared to the experimentally observed motion of tracer beads without the need of any post-processing. This allows the ALE formulation to capture the Lagrangian velocity field more accurately since in an Eulerian formulation, the Lagrangian velocity is calculated via  a post-processing step involving the computation of the Stokes drift, which depends on the gradients of the first-order velocity. The latter are difficult to capture accurately in the thin boundary layers observed in microacoustofluidic devices. Moreover, the boundary conditions employed in the ALE formulation at the actuated boundary are \emph{exact}, unlike the current approaches that utilize a second-order Taylor series expansion to obtain the boundary conditions for the acoustic streaming problem. Thus, the formulation of the second-order problem directly in terms of Lagrangian flow velocity removes the ambiguity concerning the second-order boundary condition at the oscillating walls.
 
The rest of the article is organized as follows: In Section~\ref{sec: Formulation}, we present the \ac{ALE} formulation and describe the various notions of mean fields. In Section~\ref{sec: TimescaleSeparation} we perform the separation of time-scales for microacoustofluidic devices to obtain the zeroth-, the first-, and the second-order problem. In Section~\ref{sec: NumScheme}, we describe the numerical scheme employed to solve the governing equations. Section~\ref{sec: results} presents the numerical test cases for varying actuation functions and device sizes compared to the wavelength of the acoustic actuation employed. Finally, we end the article by a brief discussion of the results in Section ~\ref{sec: Discussion}.

\section{Formulation}
\label{sec: Formulation}

\subsection{Basic notation and kinematics}
\begin{figure}[htb]
\centering\includegraphics{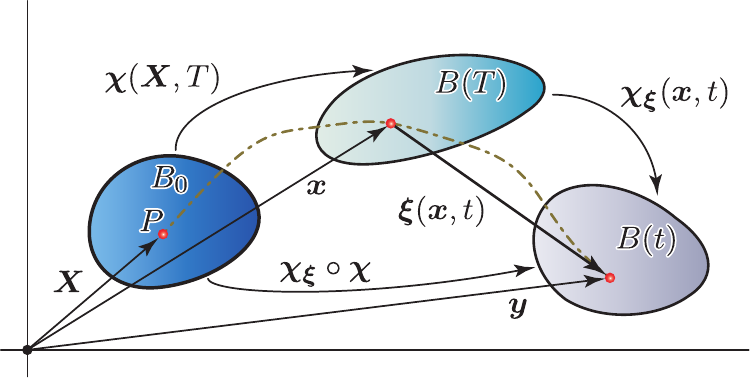}
\caption{Schematic of the domains we consider.  $B_{0}$ denotes the reference configuration of our system.  $B(t)$ denotes the current configuration of the system.  We view $B(t)$ as the outcome of a deformation $\bv{\chi}_{\bv{\xi}}$ from an intermediate configuration $B(T)$, which we view as the configuration of the body whose motion is observed at the slow time scale.  The dashed line represents the trajectory of a material particle $P$ (shown in the reference configuration).}
\label{fig: schematic}
\end{figure}

We begin by distinguishing two distinct time scales with corresponding time variables $t$ and $T$. We will refer to $t$ as the fast time characterizing the acoustic wave time scale. We will refer to $T$ as the slow time characterizing the mean motion of the fluid.  A distinction between these two times can be done by adopting the following relation \citep{Xie2014multiscale}:
\begin{equation}
\label{eq: timescales}
T=\epsilon^{\beta} t,
\end{equation}
where $\epsilon > 0$ is a nondimensional smallness parameter typically defined as a ratio of length scales \citep[cf.][]{Koster2006Numerical-Simul0,Ding2013}, and $\beta > 1$ is a nondimensional exponent, yet to be determined, whose value sets the meaning of time scale separation in a rigorous way.  Next, referring to Fig.~\ref{fig: schematic}, 
we denote by $B_{0} \subset \mathscr{E}^{d}$, $d = 2,3$, the reference configuration of the fluid, where $\mathscr{E}^{d}$ is the underlying $d$-dimensional Euclidian point space in which the body moves.  For simplicity $B_{0}$ will be taken as the initial configuration of the fluid (before acoustic excitation), assumed to be a quiescient state. We denote by $\mathscr{V}$ the translation (vector) space of $\mathscr{E}^{d}$.  As is typical in an Euclidean context, we define a second-order tensor to be an element of $\mathscr{L}(\mathscr{V},\mathscr{V})$ \citep{GurtinFried_2010_The-Mechanics_0}, i.e., a linear map from $\mathscr{V}$ into $\mathscr{V}$.

As explained later, the system's motion will be viewed as a sum of contributions of different orders.  Given a quantity, say $\phi$, the contribution of $n$-th order to this quantity will be denoted as $\order{\phi}{n}$.  Finally, given a tensor $\tensor{A}$, its transpose and its inverse transpose will be denoted by $\transpose{\tensor{A}}$ and $\inversetranspose{\tensor{A}}$, respectively.

We denote material fluid particles in $B_{0}$ by $\bv{X}$. The current configuration of the fluid resulting from the acoustic excitation will be described as follows:
\begin{equation}
\label{eq: actual motion}
\bv{\zeta}: B_{0} \to B(t) \subset\mathscr{E}^{d}, \quad 
B(t) \ni \bv{y} = \bv{\zeta}(\bv{X},t),
\end{equation}
where, as is standard in continuum mechanics, for all $t > 0$, $\bv{\zeta}(\bv{X},\bullet)$ is assumed to be a diffeomorphism between $B_{0}$ and $B(t)$, and where, for all $\bv{X} \in B_{0}$, $\bv{\zeta}(\bullet,t) \in C^{2}(\mathbb{R}^{+})$.  We assume that the motion $\bv{\zeta}(\bv{X},t)$ is the composition of two motions, with smoothness properties in space and time so as to guarantee the assumed properties of $\bv{\zeta}(\bv{X},t)$.  These motions are as follows:
\begin{alignat}{2}
\label{eq: motion composition}
\bv{\chi}&: B_{0} \to B(T) \subset\mathscr{E}^{d}, \quad& 
B(T) \ni \bv{x} &\coloneqq \bv{\chi}(\bv{X},T)
\shortintertext{and}
\bv{\chi}_{\bv{\xi}}&: B(T) \to B(t) \subset\mathscr{E}^{d}, \quad&
B(t) \ni \bv{y} &\coloneqq \bv{\chi}_{\bv{\xi}}(\bv{x},t),
\end{alignat}
such that, under the constraint in Eq.~\eqref{eq: timescales}, 
\begin{equation}
\label{Eq: totalmotion}
\bv{\zeta} = \bv{\chi}_{\bv{\xi}}\circ\bv{\chi}
\quad\text{that is}\quad
\bv{y}=\bv{\chi_{\xi}} \bigl(\bv{\chi}(\bv{X},T),t\bigr)\big|_{T = \epsilon^{\beta} t}.
\end{equation}
$\bv{\chi}(\bv{X},T)$ and $\bv{\chi}_{\bv{\xi}}(\bv{x},t)$ will be referred to as the slow and the fast motion, respectively.  Since we have defined $B_{0}$, $B(T)$, and $B(t)$ as subsets of the same underlying Euclidean point space, we introduce corresponding displacement fields as follows:
\begin{alignat}{3}
\label{eq: displacements defs 1}
&\order{\bv{u}}{0} : B_{0} \to \mathscr{V}, 
~&\order{\bv{u}}{0}(\bv{X},T) &\coloneqq \bv{\chi}(\bv{X},T) - \bv{X},
&\text{$T > 0$ and $\bv{X} \in B_{0}$},
\\
\label{eq: displacements defs 2}
&\bv{\xi} : B(T) \to \mathscr{V}, 
~&\bv{\xi}(\bv{x},t) &\coloneqq \bv{\chi}_{\bv{\xi}}(\bv{x},t) - \bv{x},
&\text{$t > 0$, $\bv{x} \in B(T)$, and $T  = \epsilon^{\beta} t$},
\\
\label{eq: displacements defs 3}
&\bv{u} : B(t) \to \mathscr{V}, 
~&\bv{u}(\bv{y},t)|_{\bv{y} = \bv{\zeta}(\bv{X},t)} &\coloneqq \bv{\zeta}(\bv{X},t) - \bv{X},
&\text{$t > 0$ and $\bv{X} \in B_{0}$,}
\end{alignat}
where, as discussed later, $\order{\bv{u}}{0}$ is the zeroth-order contribution to $\bv{u}$.  In view of Eqs.~\eqref{eq: timescales} and~\eqref{Eq: totalmotion}, we have that, for $t>0$ and $\bv{X} \in B_{0}$,
\begin{equation} 
\label{Eq: totaldisp}
\bv{u}(\bv{y},t)|_{\bv{y}=\bv{\zeta}(\bv{X},t)} = \bigl[\order{\bv{u}}{0}(\bv{X},T) + \bv{\xi}(\bv{x},t)|_{\bv{x}=\bv{\chi}(\bv{X},T)} \bigr]_{T = \epsilon^{\beta}t},
\end{equation}
The displacement field $\bv{\xi}(\bv{x},t)$ is assumed to be a harmonic function of $t$, with period equal to that of the harmonic excitation of the system.

To make the notation less cumbersome, in the remainder of the paper we will work under the conditions underlying the composition in Eq.~\eqref{Eq: totalmotion}.  Furthermore, by the notation $\partial_{\phi}(\bullet)$ we denote the partial derivative of the quantity $(\bullet)$ with respect to the variable $\phi$, where the latter can be scalar-, vector-, or tensor-valued.  With the notation $\nabla_{\bv{z}} (\bullet)$ we denote the spatial gradient of $(\bullet)$ with respect to the position variable $\bv{z}$, i.e., $\nabla_{\bv{z}} (\bullet) = \partial_{\bv{z}}(\bullet)$; and by $\nabla_{\bv{z}} \scdot (\bullet)$ we denote the corresponding divergence operator.  Finally, the notation $\dot{\phi}$ will denote the \emph{material time derivative} of the quantity $\phi$, where the latter can be scalar-, vector-, or tensor-valued, and where by \emph{material time derivative} we mean \citep[cf.][]{GurtinFried_2010_The-Mechanics_0},
\begin{equation}
\label{eq: material time derivative def}
\dot \phi \coloneqq \partial_{t} \phi|_{\text{holding $\bv{X} \in B_{0}$ fixed}}.
\end{equation}

The deformation gradient of the slow motion is
\begin{equation}
\label{eq: F0 def}
\order{\tensor{F}}{0} : B_{0} \to \mathscr{L}^{+}(\mathscr{V},\mathscr{V}), \quad
\order{\tensor{F}}{0} :=\tensor{I}+\nabla_{\bv{X}}\order{\bv{u}}{0}.
\end{equation}
Similary, the deformation gradient of the fast (harmonic) motion is 
\begin{equation}
\label{eq: Fxi def}
\tensor{F}_{\bv{\xi}} : B(T) \to \mathscr{L}^{+}(\mathscr{V},\mathscr{V}), \quad
\tensor{F}_{\bv{\xi}}:=\tensor{I}+\nabla_{\bv{x}}\bv{\xi},
\end{equation}
where $\tensor{I}$ is the identity tensor and $\mathscr{L}^{+}(\mathscr{V},\mathscr{V})$ is the set of tensors with positive determinant.  The positivity of $\det(\order{\tensor{F}}{0})$  and $\det(\tensor{F}_{\bv{\xi}})$ is a consequence of having assumed $\bv{\chi}_{\bv{\xi}}$ to be a diffeomorphism for all $T > 0$ and that $B_{0} \equiv B(0)$.  The deformation gradient of the overall motion $\bv{\zeta}$ is then given as \citep{GurtinFried_2010_The-Mechanics_0}
\begin{equation}
\label{eq: F total}
\tensor{F} : B_{0} \to \mathscr{L}^{+}(\mathscr{V},\mathscr{V}), \quad
\tensor{F} = \tensor{F}_{\bv{\xi}} \order{\tensor{F}}{0}.
\end{equation}
For future use, we will denote $\det(\tensor{F}_{\bv{\xi}})$ by $J_{\bv{\xi}}$, and by recalling the formula for the characteristic polynomial of a second-order tensor, we have the following (exact) representation formula for $J_{\bv{\xi}}$:
\begin{equation}
\label{eq: det F expansion}
J_{\bv{\xi}} = 1
+
\mathscr{I}_{1}(\nabla_{\bv{x}}\bv{\xi})
+
\mathscr{I}_{2}(\nabla_{\bv{x}}\bv{\xi})
+
\mathscr{I}_{3}(\nabla_{\bv{x}}\bv{\xi}),
\end{equation}
where $\mathscr{I}_{k}$, $k = 1, \ldots, 3$, are the principal invariants of $\nabla_{\bv{x}}\bv{\xi}$, namely
\begin{equation}
\label{eq: invariants of nabla xi}
\mathscr{I}_{1}(\nabla_{\bv{x}}\bv{\xi})
= \nabla_{\bv{x}}\scdot\bv{\xi},\quad
\mathscr{I}_{2}(\nabla_{\bv{x}}\bv{\xi})
=\tfrac{1}{2}
\bigl[
(\nabla_{\bv{x}}\scdot\bv{\xi})^{2}
-
\transpose{\nabla_{\bv{x}} \bv{\xi}} \colondot \nabla_{\bv{x}}\bv{\xi}
\bigr],\quad
\mathscr{I}_{3}(\nabla_{\bv{x}}\bv{\xi}) = \det{\nabla_{\bv{x}}\bv{\xi}},
\end{equation}

Using Eq.~\eqref{Eq: totaldisp}, the material velocity field is
\begin{equation} 
\label{eq: Eulerian v def}
\bv{v} : B(t) \to \mathscr{V}, \quad
\bv{v} \coloneqq \dot{\bv{u}}
\quad\Rightarrow\quad
\bv{v} = \partial_{t}\bv{\xi} + \epsilon^{\beta} \tensor{F}_{\bv{\xi}}\partial_{T}\order{\bv{u}}{0},
\end{equation}
where, as implied by the first of Eqs.~\eqref{eq: Eulerian v def}, $\bv{v}$ is a field over $B(t)$. For later use we define the following velocity field:
\begin{equation}
\label{eq: vxi def}
\bv{v}_{\bv{\xi}} : B(T) \to \mathscr{V},\quad
\bv{v}_{\bv{\xi}} \coloneqq \partial_{t} \bv{\xi}.
\end{equation}

\subsection{Governing Equations}
The motion of the fluid is governed by the balance laws for mass and momentum \citep{GurtinFried_2010_The-Mechanics_0} given by, respectively,
\begin{alignat}{2}
\label{Eq: MassBalance}
\dot{\rho} + \rho \nabla_{\bv{y}} \scdot \bv{v} & = 0, \quad& &\bv{y}\in B(t),
\\
\label{Eq: MomentumBalance}
\rho ( \dot{\bv{v}} - \bv{b}) - \nabla_{\bv{y}} \scdot \tensor{T} &= \bv{0}, \quad&  &\bv{y}\in B(t),
\end{alignat}
where $\rho(\bv{y},t): B(t) \to \mathbb{R}$ is the mass density distribution, $\bv{b}(\bv{y},t) : B(t) \to \mathscr{V}$ is the external body force density per unit mass, and $\tensor{T}(\bv{y},t) : B(t) \to \Lin$ is the Cauchy stress.
 
We assume the fluid to be linear, viscous, and compressible with constitutive response function for the Cauchy stress given by
\begin{equation}
\label{Eq: constitutive stress}
\tensor{T} = -p(\rho) \tensor{I} + \mu (\nabla_{\bv{y}} \bv{v} + \transpose{\nabla_{\bv{y}}\bv{v}})+\mu_{\mathrm{b}} (\nabla_{\bv{y}} \scdot \bv{v}) \tensor{I} , 
\end{equation}
where $p$ is the fluid (static) pressure, $\mu$ and $\mu_\mathrm{b}$ are constants representing the shear and bulk viscosities, respectively, and $p(\rho)$ is assumed to be the following linear relation
\begin{equation}
\label{Eq: constitutive pressure}
p = c_{0}^{2} \, (\rho - \rho_{0}), 
\end{equation}
where $c_{0}$ and $\rho_{0}$ are constants denoting the fluid's speed of sound and mass density at rest, respectively.

\subsection{Reformulation of the governing equations}
We now reformulate the governing equations by mapping them onto body's mean deformed configuration $B(T)$.  A field defined over $B(t)$ mapped onto $B(T)$ will be identified by an asterisk, i.e., given $\phi(\bv{y},t) : B(t) \to \mathscr{W}$, with $\phi$ a scalar-, vector-, or tensor-valued quantity, and $\mathscr{W}$ some appropriate vector space, $\phi^{*}(\bv{x},t) : B(T) \to \mathscr{W}$ will denote the corresponding mapped quantity according to the following definition:
\begin{equation}
\label{eq: *notation}
\phi^{*}(\bv{x},t) = \phi(\bv{y},t)|_{\bv{y} = \bv{x} + \bv{\xi}(\bv{x},t)}.
\end{equation}
A field defined over $B_{0}$ mapped onto $B(T)$ will be denoted with the superscript symbol $\#$, i.e., given $\phi(\bv{X},t) : B_{0} \to \mathscr{W}$, with $\phi$ a scalar-, vector-, or tensor-valued quantity, and $\mathscr{W}$ some appropriate vector space, $\PF{\phi}(\bv{x},t) : B(T) \to \mathscr{W}$ will denote the corresponding mapped quantity according to the following definition:
\begin{equation}
\label{eq: hash notation}
\PF{\phi}(\bv{x},t) = \phi(\bv{X},t)|_{\bv{X} = \bv{\chi}^{-1}(\bv{x},t)}.
\end{equation}

Using the above definition, Eqs.~\eqref{Eq: MassBalance} and~\eqref{Eq: MomentumBalance} can be expressed on $B(T)$ as (see Appendix~A for details)
\begin{alignat}{2}
\label{Eq: mappedmassbalance}
\partial_{t}\rho^{*} + \inversetranspose{\tensor{F}}_{\bv{\xi}} \nabla_{\bv{x}}\rho^{*}\cdot(\bv{v}^{*}-\bv{v}_{\bv{\xi}}) +\rho^{*} \inversetranspose{\tensor{F}}_{\bv{\xi}} \colondot \nabla_{\bv{x}}\bv{v}^{*} &= 0, \quad& &\text{in $B(T)$},
\\
\label{Eq: mappedmomentumstrong}
J_{\bv{\xi}} \rho^{*}
\bigl[\partial_{t}\bv{v}^{*} + \nabla_{\bv{x}} \bv{v}^{*} \tensor{F}_{\bv{\xi}}^{-1} (\bv{v}^{*} - \bv{v}_{\bv{\xi}})- \bv{b}^{*}\bigr] 
- \nabla_{\bv{x}} \cdot \tensor{P}^{*} &= \bv{0}, \quad& &\text{in $B(T)$},
\end{alignat}
where, with a slight abuse of notation, $\tensor{P}^{*}$ is the Piola-Kirchhoff stress tensor on $B(T)$ \citep[cf.][]{GurtinFried_2010_The-Mechanics_0} defined as follows:
\begin{equation}
\label{eq: P* def}
\tensor{P}^{*} : B(T) \to \Lin,\quad
\tensor{P}^{*} \coloneqq J_{\xi} \tensor{T}^{*} \inversetranspose{\tensor{F}}_{\bv{\xi}}.
\end{equation}
For later use, it is also convenient to consider the Lagrangian expression of the balance of mass.  This reads $\rho(\bv{y},t) \det{\tensor{F}(\bv{X},t)} = \rho_{0}(\bv{X})$ with $\bv{y} = \bv{\zeta}(\bv{X},t)$ \citep{GurtinFried_2010_The-Mechanics_0}.  Mapping this expression onto $B(T)$, and in view of Eqs.~\eqref{eq: F total} and~\eqref{eq: hash notation}, we have
\begin{equation}
\label{eq: Lagrangian mass balance}
\rho^{*} \det(\tensor{F}_{\bv{\xi}}) \det(\PFF) = \PF{\rho}_{0}.
\end{equation}

\subsection{Time Average, Time Scale Separation, and Orders of Magnitude}
\label{sec: time averaging and more}
As anticipated earlier, to model the behavior of acoustofluidic devices, we assume that the displacement field $\bv{\xi}(\bv{x},t)$ is the time-harmonic response to the time-harmonic excitation of the system. We will denote the period of this excitation by $\Pi$.

We now introduce the notation to denote the time average operation over the excitation period $\Pi$.  For any generic quantity $\phi(\bullet, t)$ (scalar-, vector-, or tensor-valued), the notation $\langle \phi \rangle$ will identify its time average over the period of harmonic excitation, i.e.,
\begin{equation}
\label{eq: Pi avg def}
\langle \phi \rangle(\bullet, t) \coloneqq \frac{1}{\Pi} \int_{t}^{t+\Pi} \phi(\bullet, \tau) \, \d{\tau},
\end{equation}
We note that, having assumed that $\bv{\xi}(\bv{x},t)$ is a harmonic function of time with period $\Pi$, we have
\begin{equation}
\label{eq: xi avg def}
\langle \bv{\xi} \rangle = 0.
\end{equation}
A similar conclusion applies to any $\Pi$-periodic harmonic function.  On the other hand, we observe that $\tensor{F}_{\bv{\xi}}$ is merely $\Pi$-periodic and therefore $\langle \tensor{F}_{\bv{\xi}} \rangle$ does not vanish.  In fact, from the last of Eqs.~\eqref{eq: Fxi def}
\begin{equation}
\label{eq: Fxi avg}
\langle \tensor{F}_{\bv{\xi}} \rangle = \tensor{I}.
\end{equation}
For future discussion, it is also important to be able to relate $\langle \phi \rangle$ for a generic function $\phi(\bullet, t)$ to the smallness parameter $\epsilon$.  To this end, referring to Eq.~\eqref{eq: timescales}, we set
\begin{equation}
\label{eq: time pull back}
\bar{\phi}(\bullet,T) \coloneqq \phi(\bullet,t)|_{t = \epsilon^{-\beta}T}.
\end{equation}
The above expression is a simple re-expression of a quantity from the fast to the slow time scale. Next, assuming that $\phi(\bullet,t)$ is differentiable in time, in the neighborhood of a time instant $t = t_{0}$, Taylor's theorem allows us to write
\begin{equation}
\label{eq: Taylor Th}
\phi(\bullet,t) = \phi(\bullet,t_{0}) + \partial_{t}\phi(\bullet,t_{0}) (t - t_{0}) + o\Bigl(|t - t_{0}|\Bigr).
\end{equation}
Therefore, letting $T_{0} = \epsilon^{\beta} t_{0}$, for such a $\phi(\bullet,t)$, we can express $\langle \phi \rangle(\bullet,t)$ in a neighborhood of $t_{0}$ as
\begin{equation}
\label{eq: Taylor Th explicit}
\langle \phi \rangle(\bullet,t) = \bar{\phi}(\bullet,T_{0}) + \tfrac{1}{2}\epsilon^{\beta}\Pi \partial_{T}\bar{\phi}(\bullet,T_{0}) + \epsilon^{\beta}\partial_{T}\bar{\phi}(\bullet,T_{0}) (t - t_{0}) + o\Bigl(|t - t_{0}|^{2}\Bigr).
\end{equation}
Next, we observe that
\begin{equation}
\label{eq: Taylor Th explicit transition}
\epsilon^{\beta}(t - t_{0}) = T - T_{0}
\quad\text{and}\quad
\bar{\phi}(\bullet,T) = \bar{\phi}(\bullet,T_{0}) + \partial_{T}\bar{\phi}(\bullet,T_{0}) (T - T_{0}) + \epsilon^{\beta} o\Bigl(|t - t_{0}|\Bigr).
\end{equation}
Now, letting $\Delta T = \mathcal{O}(1)$ be some meaningful time duration at the slow scale, in the study of microacoustofluidics it is reasonable to posit that the excitation period is a reference time interval for distinguishing fast motions from slow motions.  Furthermore, in view of the definition in Eq.~\eqref{eq: Pi avg def}, it is reasonable to limit the context of the expansion in Eq.~\eqref{eq: Taylor Th explicit} to sub-intervals of the period $\Pi$. Therefore we posit that
\begin{equation}
\label{eq: period scale}
\text{$\Pi = \epsilon^{\gamma} \Delta T$ with $\gamma \geq \beta$ and $|t - t_{0}| < \Pi$.}
\end{equation}
With this stipulation, Eqs.~\eqref{eq: Taylor Th explicit} and~\eqref{eq: Taylor Th explicit transition} imply that, at worst,
\begin{equation}
\label{eq: Taylor Th explicit final}
\langle \phi \rangle(\bullet,t) = \bar{\phi}(\bullet,T) + \mathcal{O}\bigl(\epsilon^{2 \beta}\bigr),\end{equation}
that is, the time average operation is, at worst, equivalent to the remapping at the slow scale up to a remainder on the order of $\epsilon^{2\beta}$.

As indicated in the Introduction, we will follow the approach in \citet{Nyborg1998Acoustic-Stream0}  \citep[see also][]{koster2007numerical,muller2012numerical,nama2015numerical,Vanneste2011Streaming-by-Le0} where the response to a harmonic excitation is analyzed via an asymptotic expansion of the velocity in contributions of different integer orders of the smallness parameter $\epsilon$.  Our approach differs from previous ones in that this expansion is done on the equations written over $B(T)$ instead of $B(t)$.  Furthermore, as done in \citet{Xie2014multiscale}, we add the displacement to the fields to be expanded, with the understanding that the displacement field must be allowed to be on the order of the device's largest linear dimension. Hence, we set the displacement field describing the mean motion of the system to be of order $\mathcal{O}(\epsilon^{0})=\mathcal{O}(1)$, and we have already denoted this field by $\order{\bv{u}}{0}$.  The remaining contribution to the displacement is then the harmonic contribution $\bv{\xi}$, which we assume to be $\mathcal{O}(\epsilon)$. Also, as it is customary, we will assume that differentiation with respect to any of the position variables and with respect to the fast time $t$ does not alter the order of a quantity. Here, we remark that we are following in the footsteps of analyses like that in \citet{koster2007numerical}.  In reality, this is not the only way to carry out asymptotic expansions, as shown in \citet{Xie2014multiscale} in which very interesting alternative scenarios have been identified.  The current work will explore such scenarios in future publications.

Summarizing, we have
\begin{equation}
\label{eq: order of magnitude}
\order{\bv{u}}{0} = \mathcal{O}(1),\quad
\order{\tensor{F}}{0} = \mathcal{O}(1),\quad
\bv{\xi} = \mathcal{O}(\epsilon),\quad
\nabla_{\bv{x}}\bv{\xi} = \mathcal{O}(\epsilon),\quad
\bv{v}_{\bv{\xi}} = \mathcal{O}(\epsilon),\quad
\tensor{F}_{\bv{\xi}} = \mathcal{O}(1).
\end{equation}
With reference to Eq.~\eqref{eq: det F expansion}, we also have
\begin{equation}
\label{eq: order of invariants}
\mathscr{I}_{1}(\nabla_{\bv{x}}\bv{\xi}) = \mathcal{O}(\epsilon),
\quad
\mathscr{I}_{2}(\nabla_{\bv{x}}\bv{\xi}) = \mathcal{O}\bigl(\epsilon^{2}\bigr),
\quad
\mathscr{I}_{3}(\nabla_{\bv{x}}\bv{\xi}) = \mathcal{O}\bigl(\epsilon^{3}\bigr).
\end{equation}
We will then follow previous approaches \citep{koster2007numerical,muller2012numerical,nama2015numerical,Vanneste2011Streaming-by-Le0} in assuming that
\begin{enumerate}[(i)]
\item
 there is no order zero (background) velocity field;
\item
 the harmonic response in terms of velocity is of order $\epsilon$;
\item 
and the streaming flow is at most of order $\epsilon^{2}$.
\end{enumerate}
Using these assumptions, and enforcing consistency between the material time derivative of the displacement expansion and that of the velocity yields a constraint equation for $\beta$ that determines the separation of time scales.  Specifically, referring to Eq.~\eqref{eq: Eulerian v def}, the mapped material velocity is given by
\begin{equation}
\label{eq: material v B(T)}
\bv{v}^{*}
=
\bv{v}_{\bv{\xi}}
+
\epsilon^{\beta}\tensor{F_{\xi}} \partial_{T}\order{\bv{u}}{0}.
\end{equation}
In view of Eqs.~\eqref{eq: order of magnitude}, for the first-order contribution to be only $\bv{v}_{\bv{\xi}}$, the last term must necessarily be $\mathcal{O}(\epsilon^{2})$.  This result matches the analysis of the ``non-inertial regime'' described by~\citet{Xie2014multiscale}. We then formally set
\begin{equation}
\label{eq: beta choice}
\beta = 2,
\end{equation}
for which the first-order contribution to the material velocity is
\begin{equation}
\label{eq: v orders 1 def}
\order[*]{\bv{v}}{1}(\bv{x},t) = \bv{v}_{\bv{\xi}}(\bv{x},t),
\end{equation}
whereas the second-order contribution is defined as follows:
\begin{equation}
\label{eq: v orders 2 def}
\order[*]{\bv{v}}{2} : B(T) \to \mathscr{V}, \quad
\order[*]{\bv{v}}{2}(\bv{x},t) \coloneqq 
\epsilon^{2}\tensor{F_{\bv{\xi}}}(\bv{x},t)\Bigl(\partial_{T}\order{\bv{u}}{0}(\bv{X},T)\Big|_{\bv{X}=\chi^{-1}(\bv{x},T),\,T=\epsilon^{2}t}\Bigr)
\end{equation}

As far as the mass density distribution is concerned, we simply postulate the existence of the zeroth-, first-, and the second-order contributions.  With this said, we make the additional assumption that the time scale at which the second-order contribution is meaningful is the slow time scale.  In other words, we posit that
\begin{equation}
\label{eq: order assumption for density}
\rho^*(\bv{x},t) = \order[*]{\rho}{0}(\bv{x},t)+\order[*]{\rho}{1}(\bv{x},t) + \order[*]{\rho}{2}(\bv{x},t).
\end{equation}
Finally, we assume that the mass density distributions and external body force field are as follows:
\begin{equation}
\label{eq: orders of magnitude for density and body force}
\rho_{0} = \mathcal{O}(1)
\quad\text{and}\quad
\bv{b}^{*} = \mathcal{O}(1).
\end{equation}

\subsection{Mean Fields and Lagrangian-mean Fields}
Before presenting the hierarchy of \ac{BVPs} generated by the assumptions stated in the previous section, we now revisit our kinematics and, discuss the rationale for our assumptions.

In references \citet{Nyborg1998Acoustic-Stream0,koster2007numerical,muller2012numerical,nama2015numerical,Vanneste2011Streaming-by-Le0,Bradley1996Acoustic-Stream0,bradley2012acoustic} (and the many works that adopt the same kinematic framework) the domain of the equations governing the response of the fluid under a harmonic excitation, is a \emph{time-independent} domain (or control volume), which we will call $\Omega$. The fields defined over $\Omega$ are expressed as functions of time $t$ and a position variable, which we will denote by $\bv{z}$.  The purpose of the present discussion is to compare and contrast the kinematics adopted in the paper with that of previous works.

We begin by comparing the meaning of the variable $\bv{z}$ with the three descriptors used here, namely, $\bv{X}$, $\bv{x}$, and $\bv{y}$. Similar to what we have done, in the aforementioned references there is a slow and a fast time scale, and quantities at the slow time scale are averages of analogous quantities at the fast time scale.  In the \ac{GLM} theory \citep{Andrews1978An-Exact-Theory-0,Buhler2009Waves-and-Mean-0} this averaging concept is also applied to the position $\bv{z}$ itself.  Let's recall that $\bv{z}$ is viewed as the position at time $t$ of some material particle $P$, with the stipulation that at time $t + \Delta t > t$ the position $\bv{z}$ will be occupied by a particle $Q$ such that, in general, $Q \ne P$. If this understanding of $\bv{z}$ is the only one used, then $\bv{z}$ does not have a clear correspondence with any of the three descriptors introduced here.  The situation is different if we interpret $\bv{z}$ as is done in the \ac{GLM} theory, which posits that $\bv{z}$ is the \emph{mean} position of $P$ in the sense that $\bv{z}$ is the average of all the positions that $P$ occupies over the period of excitation surrounding the time instant $t$.  It is this idea that is then developed into the notions like Stokes drift and Lagrangian-mean motion. To be explicit, as discussed in, say, references \citet{Andrews1978An-Exact-Theory-0,Buhler2009Waves-and-Mean-0}, a particle with nominal position $\bv{z}$ at time $t$ has true position at $\bv{z} + \bv{q}(\bv{z},t)$. This is formalized as a map (cf.\, e.g., Eq.~(2.2) in \citet{Andrews1978An-Exact-Theory-0}) as follows:
\begin{equation}
\label{eq: LM mapping def}
\bv{z} \mapsto \bv{z} + \bv{q}(\bv{z},t),
\end{equation}
where $\bv{q}$ is referred to as the ``disturbance-associated particle displacement field'' in \citet{Andrews1978An-Exact-Theory-0}, and as the ``lift field'' in \citet{Buhler2009Waves-and-Mean-0}, and where $\bv{q}(\bv{z},t)$ is required to have the following property:
\begin{equation}
\label{eq: lift field property}
\langle \bv{q} \rangle = \bv{0}.
\end{equation}
This map is then used to define the concept of Lagrangian-mean average.  In this paper, the  Lagrangian-mean of a field $\phi(\bv{z},t)$ will be denoted by $\LM{\phi}(\bv{z},t)$ and defined as in \citet{Andrews1978An-Exact-Theory-0,Buhler2009Waves-and-Mean-0}, i.e., 
\begin{equation}
\label{eq: LM Def}
\LM{\phi}(\bv{z},t) \coloneqq \langle\phi(\bv{z} + \bv{q}(\bv{z},t),t)\rangle.
\end{equation}
Applying Eqs.~\eqref{eq: lift field property} and~\eqref{eq: LM Def}
\begin{equation}
\label{eq: LM consistency}
\LM{\bv{z}} =  \langle\bv{z} + \bv{q}(\bv{z},t) \rangle
\quad\Rightarrow\quad
\LM{\bv{z}} = \bv{z},
\end{equation}
so that the nominal position $\bv{z}$ is (by definition) the Lagrangian-mean of the true position.  Focusing now on our kinematics, we recall it is the position $\bv{y}$ that identifies points in the current configuration. Hence, using Eqs.~\eqref{eq: actual motion} and~\eqref{eq: displacements defs 2}, we have
\begin{equation}
\label{eq: LM map}
\bv{y} = \bv{x} + \bv{\xi}(\bv{x},t)
\quad\Rightarrow\quad
\langle \bv{y} \rangle = \bv{x}.
\end{equation}
Therefore, the kinematic framework we have adopted is simply that of the standard \ac{GLM} theory, where $\bv{x}$ is the Lagrangian-mean position and $\bv{\xi}(\bv{x},t)$ is the lift field.  This is not surprising, in that we have adopted an \ac{ALE} perspective and the  \ac{ALE} mapping from the current domain to a convenient (e.g., computationally) domain $B(T)$ is exactly the map in Eqs.~\eqref{eq: LM mapping def} in the \ac{GLM} theory.  That is, at least when it comes to the kinematics of position, the \ac{GLM} theory is an \ac{ALE} scheme.  At this point it is important to remark that there is no difficulty in reconciling that the domain $B(T)$ is time dependent whereas $\Omega$ is not, even in the \ac{GLM} theory.  As already remarked, $\Omega$ is time independent because it is a control volume. By contrast $B(T)$ is a configuration and as such it is time dependent because material particles have moved to go from $B_{0}$ to $B(T)$.  However, if the particles on the boundary of $B(T)$ do not move, i.e., if $\order{\bv{u}}{0}(\bv{X},T) = \bv{0}$ for $\bv{X} \in \partial B_{0}$, then $B(T)$ will always occupy the same portion of $\mathscr{E}^{d}$ as $B_{0}$, and it is this portion that coincides with $\Omega$.  In other words, $\Omega$ and $B(T)$ coincide if they are both viewed as mere (geometric) point sets.  We contend that, in the study of acoustofluidic devices, this boundary condition is a reasonable and convenient one to assume, namely, that the motion of the system, on average, does not displace the boundary of the fluid compartment of the device.  Hence, in the remainder of the paper we formally assume that
\begin{equation}
\label{eq: u0 boundary condition}
\order{\bv{u}}{0}(\bv{X},T) = \bv{0},\quad \bv{X} \in \partial B_{0}
\quad\Rightarrow\quad
\partial B_{0} \equiv \partial B(T).
\end{equation}

We now turn to comparing the velocity descriptors.   We start by observing that the velocity field $\bv{v}(\bv{y},t)$ does not find correspondence in the standard approaches because its domain is $B(t)$, which is a time-dependent deformable domain, as opposed to the control volume $\Omega$.  However, as we have already remarked that our kinematic framework is that of the \ac{GLM} theory, we should be able to find correspondences between said theory and our approach.  To this end, let's recall that
\begin{equation}
\label{eq: LLM v def}
\LM{\bv{v}}(\bv{x},t) = \langle\bv{v}(\bv{y},t)\rangle
\quad\Rightarrow\quad
\LM{\bv{v}}(\bv{x},t) = \langle\bv{v}\rangle(\bv{y},t)|_{\bv{y} = \bv{x} + \bv{\xi}(\bv{x},t)}.
\end{equation}
Hence, from Eq.~\eqref{eq: Eulerian v def}, and recalling that $\partial_{t}\bv{\xi}$ is harmonic and $\epsilon^{\beta} \partial_{T}\order{\bv{u}}{0} =  \partial_{t}\order{\bv{u}}{0}$, we have
\begin{equation}
\label{eq: LLM v  u0 rel}
\LM{\bv{v}}(\bv{x},t)
=
\langle
\tensor{F}_{\bv{\xi}} \partial_{t} \order{\bv{u}}{0}
\rangle(\bv{x},t).
\end{equation}
Referring to Eq.~\eqref{eq: v orders 2 def}, we also see that
\begin{equation}
\label{eq: LLM v v*2 rel}
\LM{\bv{v}}(\bv{x},t)
=
\langle
\order[*]{\bv{v}}{2}
\rangle(\bv{x},t).
\end{equation}
Both Eq.~\eqref{eq: LLM v  u0 rel} and~\eqref{eq: LLM v v*2 rel} are results worth discussing.  Looking ahead at the numerical implementation of our scheme, we will select $\langle\order[*]{\bv{v}}{2}\rangle$ as one of the primary unknowns of our time-averaged second-order problem.  Hence, Eq.~\eqref{eq: LLM v v*2 rel} implies that the Lagrangian-mean velocity is $\mathcal{O}(\epsilon^{2})$ and it is precisely one of the fields that our solution scheme delivers directly.

Another issue concerning our kinematic choices concerns the field $\order{\bv{u}}{0}$. In a fluids problem in which there is no need to determine the displacement field, one can simply not solve for it.  However, being able to solve for the field $\order{\bv{u}}{0}$, whether as a field with domain $B_{0}$ or as a field $\PFu$ defined over $B(T)$ (see definition in Eq.~\eqref{eq: hash notation}), is essential when interested in modeling the streaming flows of complex fluids that have an elastic component to their constitutive response and/or to problems in which there are deformable solid objects immersed in the  fluid.  Hence, we discuss here how one could pose the problem of determining such a field. Clearly, any such discussion needs to start from considering carefully Eq.~\eqref{eq: LLM v  u0 rel}. If the displacement field is described through the function $\PFu$, then Eq.~\eqref{eq: LLM v  u0 rel} must be rewritten as
\begin{equation}
\label{eq: LLM v  u0 rel hash}
\LM{\bv{v}}(\bv{x},t)
=
\Bigl\langle
\tensor{F}_{\bv{\xi}} \bigl(\tensor{I} - \nabla_{\bv{x}} \PFu\bigr)^{-1} \partial_{t} \PFu
\Bigr\rangle(\bv{x},t).
\end{equation}
Whether in using Eq.~\eqref{eq: LLM v  u0 rel} or~\eqref{eq: LLM v  u0 rel hash}, the presence of a time averaging operator makes the relationship between the primary unknown $\LM{\bv{v}}$ and the displacement a nonlinear integro-differential equation. In those cases where considerations based on the smallness parameter $\epsilon$ allow it, it is possible to consider an approximation that reduces to a partial differential equation.  To illustrate this point, we will focus on Eq.~\eqref{eq: LLM v  u0 rel}.  From the discussion of Eq.~\eqref{eq: v orders 2 def}, we know that $\partial_{t}\order{\bv{u}}{0} = \mathcal{O}\bigl(\epsilon^{\beta}\bigr)$.  Since $\tensor{F}_{\bv{\xi}}\partial_{t}\order{\bv{u}}{0} = \partial_{t}\order{\bv{u}}{0} + \nabla_{\bv{x}}\bv{\xi} \partial_{t}\order{\bv{u}}{0}$, in view of the third of Eqs.~\eqref{eq: order of magnitude}, we can then write
\begin{equation}
\label{eq: LLM v  u0 rel order of magnitude}
\LM{\bv{v}}
=
\partial_{t}\order{\bv{u}}{0} + \mathcal{O} \bigl(\epsilon^{\beta+1}\bigr).
\end{equation}
This result implies that if an approximation up to order $\beta+1$ is acceptable, one can reconstruct the displacement field directly from the above equation.  We note that, from a numerical viewpoint, \ac{FSI} schemes called \emph{immersed methods} \citep{Heltai2012Variational-Imp0,Peskin_2002_The-immersed_0,WangLiu_2004_Extended_0} are able to reconstruct the displacement field from an equation such as Eq.~\eqref{eq: LLM v  u0 rel order of magnitude}.  Higher order corrections that take into account the time averaging operations are possible and, in principle, can also be handled via immersed methods.

Before going to the next section, we observe that Eq.~\eqref{eq: LLM v  u0 rel order of magnitude}, along with Eq.~\eqref{eq: u0 boundary condition} implies that
\begin{equation}
\label{eq: vL boundary condition}
\LM{\bv{v}} = \bv{0}
\quad\text{on}\quad
\partial B(T).
\end{equation}

\section{Hierarchical Boundary Value Problems}
\label{sec: TimescaleSeparation}
In this section we present the hierarchy of boundary value problems that are obtained when adopting the scale separation strategy described earlier.  For each of the boundary value problem presented, we will discuss, in as much as it is reasonable, the problem's boundary conditions.  Clearly, a specific problem will justify a corresponding set of boundary conditions.  Here we limit ourselves to include the boundary conditions used later in the examples.  The latter are based on the applications of interest to our group concerning on-chip microacoustofluidic devices.

\subsection{Remarks on the balance of mass}
\label{subsec Remarks on the balance of mass}
The character of the boundary value problems induced by the selected orders of magnitude is somewhat different, to the best of the authors' knowledge, from what is currently found in the literature.  To elucidate this point, we begin with considerations based on the balance of mass in Lagrangian form reported in Eq.~\eqref{eq: Lagrangian mass balance}.  Using Eqs.~\eqref{eq: det F expansion} and~\eqref{eq: order assumption for density}, we can rewrite the balance of mass as
\begin{equation}
\label{eq: BMass orders step 1}
(\order[*]{\rho}{0} + \order[*]{\rho}{1} + \order[*]{\rho}{2})
\bigl[
1
+
\mathscr{I}_{1}(\nabla_{\bv{x}}\bv{\xi})
+
\mathscr{I}_{2}(\nabla_{\bv{x}}\bv{\xi})
+
\mathscr{I}_{3}(\nabla_{\bv{x}}\bv{\xi})
\bigr]
\det(\PFF) = \PF{\rho}_{0}.
\end{equation}
Expanding, using Eqs.~\eqref{eq: invariants of nabla xi}, and matching orders we obtain the following relations:
\begin{align}
\label{eq: BMass Lagrangian solution order 0}
\order[*]{\rho}{0} \det(\PFF) &= \PF{\rho}_{0},
\\
\label{eq: BMass Lagrangian solution order 1}
\order[*]{\rho}{1} + \order[*]{\rho}{0} \nabla_{\bv{x}} \scdot \bv{\xi} &= 0,
\\
\label{eq: BMass Lagrangian solution order 2}
\order[*]{\rho}{2}
+ \order[*]{\rho}{1} \nabla_{\bv{x}} \scdot \bv{\xi}
+ \tfrac{1}{2}\order[*]{\rho}{0} \bigl[(\nabla_{\bv{x}}\scdot\bv{\xi})^2 -
\transpose{(\nabla_{\bv{x}}\bv{\xi})} \colondot \nabla_{\bv{x}}\bv{\xi} \bigr] &= 0.
\end{align}
For the case with $\rho_{0}$ a constant, as is typically the case for fluids, we can then take the partial derivative of the above relations with respect to time while holding $\bv{x}$ fixed to obtain
\begin{align}
\label{eq: BMass Lagrangian solution order 0 rate}
\partial_{t}\order[*]{\rho}{0}  &= 0,
\\
\label{eq: BMass Lagrangian solution order 1 rate}
\partial_{t}\order[*]{\rho}{1} + \order[*]{\rho}{0} \nabla_{\bv{x}} \scdot \order[*]{\bv{v}}{1} &= 0,
\\
\label{eq: BMass Lagrangian solution order 2 rate}
\partial_{t}\order[*]{\rho}{2}
+ \order[*]{\rho}{1} \nabla_{\bv{x}} \scdot \order[*]{\bv{v}}{1}
- \order[*]{\rho}{0} \transpose{(\nabla_{\bv{x}}\bv{\xi})} \colondot \nabla_{\bv{x}}\order[*]{\bv{v}}{1} &= 0,
\end{align}
where we have used the fact that $\partial_{t} \det(\PFF) = \mathcal{O}\bigl(\epsilon^{\beta}\bigr)$.  Of the above relatioships, perhaps the most important are Eq.~\eqref{eq: BMass Lagrangian solution order 2} and~\eqref{eq: BMass Lagrangian solution order 2 rate} because they indicate that the proposed kinematic setting implies that second-order mass density distribution is completely determined by the first-order solution to the problem. Furthermore, recalling that the time average of harmonic functions over the excitation period is time independent, from Eq.~\eqref{eq: BMass Lagrangian solution order 2} we see that $\langle\order[*]{\rho}{2}\rangle$ is no longer a function of time.  In turn this implies that one can calculate the steady (slow-time) value of $\langle\order[*]{\rho}{2}\rangle$ from the first-order solution.  Later we will show that Eq.~\eqref{eq: BMass Lagrangian solution order 2 rate} allows us to obtain a corresponding result for the second-order velocity solution such that it is possible to solve directly for the steady value of  $\langle\order[*]{\bv{v}}{2}\rangle$. 

\subsection{Zeroth-order subproblem}
We now derive the zeroth-order problem by substituting the expansions of velocity and mass density and retaining only zeroth-order terms. Recalling that we have neglected the background flow, the zeroth-order subproblem is governed by the following relations:
\begin{equation}
\label{Eq: order zero problem}
\partial_{t} \order[*]{\rho}{0} = 0,\quad
\nabla_{\bv{x}} \scdot \order[*]{\tensor{P}}{0} + \order[*]{\rho}{0} \bv{b}^{*} = \bv{0}, \quad
\order[*]{\tensor{P}}{0} = -c_{0}^{2} (\order[*]{\rho}{0} - \rho_{0}) \tensor{I}.
\end{equation}
Observing that the first of Eqs.~\eqref{Eq: order zero problem} implies that $\order[*]{\rho}{0}$ is a constant, for constant $\bv{b}^{*}$, we see that the solution to the zeroth-order problem is represented by the static equilibrium state of the fluid. For convenience, we will proceed forward under the assumption that the body forces are negligible.  In this case, the solution of the zeroth-order problem is
\begin{equation}
\label{eq: 0th order solution}
\order[*]{\rho}{0} = \rho_{0}
\quad\text{and}\quad
\order[*]{\tensor{P}}{0} = \tensor{0}.
\end{equation}
The first of Eq.~\eqref{eq: 0th order solution} along with Eq.~\eqref{eq: BMass Lagrangian solution order 0} imply that 
\begin{equation}
\label{eq: volume preserving slow motion}
\det(\PFF) \approx 1,
\end{equation}
which, in turn, implies that the zeroth-order motion is expected to be volume preserving under the current assumptions.

\subsection{Acoustic subproblem}
\label{subsec: acoustic subproblem}
Next, we consider the acoustic subproblem (on the fast time scale) by substituting the expressions for velocity and mass density in terms of their components in the balance laws and retaining only the first-order terms. Using Eq.~\eqref{Eq: mappedmassbalance} and Eq.~\eqref{Eq: mappedmomentumstrong} along with the expansions of velocity and mass density, we obtain the following set of equations for the acoustic subproblem:
\begin{equation}
\label{eq: order 1 problem}
\partial_{t}\order[*]{\rho}{1} + \order[*]{\rho}{0} \nabla_{\bv{x}} \scdot \order[*]{\bv{v}}{1} = 0
\quad\text{and}\quad
\order[*]{\rho}{0} \partial_{t} \order[*]{\bv{v}}{1} - \nabla_{\bv{x}} \scdot \order[*]{\tensor{P}}{1} =\bv{0},
\end{equation}
where
\begin{equation}
\label{eq: order 1 P}
\order[*]{\tensor{P}}{1}
=
-
c_{0}^{2}
\order[*]{\rho}{1} \tensor{I}
+
\mu \Bigl(\nabla_{\bv{x}} \order[*]{\bv{v}}{1} + \nabla_{\bv{x}} \orderT[*]{\bv{v}}{1}\Bigr) + \mu_{\mathrm{b}} \bigl(\nabla_{\bv{x}} \scdot \order[*]{\bv{v}}{1}\bigr) \tensor{I}.
\end{equation}
We notice that the first of Eq.~\eqref{eq: order 1 problem} is a consequence of Eq.~\eqref{eq: BMass Lagrangian solution order 1}, the latter providing an expression of $\order[*]{\rho}{1}$ in terms of the displacement $\bv{\xi}$.  In fact, recalling that $\order[*]{\rho}{0} = \rho_{0}$, and along with Eqs.~\eqref{eq: vxi def} and~\eqref{eq: v orders 1 def}, the first-order problem takes on the following form:
\begin{equation}
\label{eq: first order problem xi}
\rho_{0} \partial_{tt} \bv{\xi} - \nabla_{\bv{x}}\scdot\order[*]{\tensor{P}}{1} = \bv{0}
\quad\text{with}\quad
\order[*]{\tensor{P}}{1}
=
\bigl(c_{0}^{2}\rho_{0}+\mu_{\mathrm{b}}\partial_{t}\bigr)(\nabla_{\bv{x}}\scdot\bv{\xi})\tensor{I}
+
\mu \partial_{t}\Bigl(\nabla_{\bv{x}} \bv{\xi} + \transpose{\nabla_{\bv{x}}\bv{\xi}}\Bigr)
\end{equation}
For the above equation we only seek harmonic solutions.  If we then let the solution be of the form $\bv{\xi}(\bv{x},t) = e^{\ii \omega t} \hat{\bv{\xi}}(\bv{x})$, with $\hat{\bv{\xi}}$ a complex-valued vector field, the first-order problem reduces to a linear elasto-statics problem with complex  moduli.

As far as the boundary conditions are concerned, there are several sets of interest in applications.  In fact, the boundary problem of interest is a coupled problem in which the acoustics is induced by an electrical actuator for which the electric field is prescribed and for which the electric actuation produces a corresponding mechanical actuation that depends on the geometry and material make-up of the device.  In this paper we do not consider the fully coupled electro-mechanical-fluidic problem. Instead we simply assume that the displacement field $\bv{\xi}$ is prescribed everywhere on the boundary:
\begin{equation}
\label{eq: BCs first order problem}
\bv{\xi}(\bv{x},t) = \bv{\xi}_{g}(\bv{x},t), \quad \bv{x} \in \partial B(T),
\end{equation}
where the subscript $g$ stands for \emph{given}.

\subsection{Mean dynamics subproblem}
Proceeding in a similar manner as done above, we derive the second-order problem. Using Eq.~\eqref{Eq: mappedmassbalance} and Eq.~\eqref{Eq: mappedmomentumstrong} along with the expansions of velocity and mass density, for the balance of mass and momentum we obtain, respectively:
\begin{align}
\label{eq: mass balance second order}
\partial_{t}\order[*]{\rho}{2}
+
\order[*]{\rho}{0} \nabla_{\bv{x}}\scdot\order[*]{\bv{v}}{2}
+
\order[*]{\rho}{1} \nabla_{\bv{x}}\scdot\order[*]{\bv{v}}{1}
-
\order[*]{\rho}{0} \transpose{(\nabla_{\bv{x}}\bv{\xi})} \colondot \nabla_{\bv{x}} \order[*]{\bv{v}}{1} & = 0
\shortintertext{and}
\label{eq: momentum balance second order}
\order[*]{\rho}{0}\partial_{t}\order[*]{\bv{v}}{2}
+
\order[*]{\rho}{1} \partial_{t}\order[*]{\bv{v}}{1}
+
\order[*]{\rho}{0} (\nabla_{\bv{x}}\scdot{\bv{\xi}})\partial_{t} \order[*]{\bv{v}}{1}
-
\nabla_{\bv{x}}\scdot \order[*]{\tensor{P}}{2}
&=
\bv{0},
\end{align}
where
\begin{equation}
\label{eq: P second order}
\begin{aligned}[b]
\order[*]{\tensor{P}}{2} &= 
-
c_{0}^{2}
\order[*]{\rho}{2} \tensor{I}
+
\mu \Bigl(\nabla_{\bv{x}} \order[*]{\bv{v}}{2} + \nabla_{\bv{x}} \orderT[*]{\bv{v}}{2}\Bigr) + \mu_{\mathrm{b}} \bigl(\nabla_{\bv{x}} \scdot \order[*]{\bv{v}}{2}\bigr) \tensor{I}
\\
&\qquad
-\mu \Bigl(\nabla_{\bv{x}} \order[*]{\bv{v}}{1} \nabla_{\bv{x}}\bv{\xi} + \transpose{\nabla_{\bv{x}}\bv{\xi}} \nabla_{\bv{x}} \orderT[*]{\bv{v}}{1}\Bigr)
-
\mu_{\mathrm{b}} \bigl( \transpose{\nabla_{\bv{x}}\bv{\xi}} \colondot \nabla_{\bv{x}}\order[*]{\bv{v}}{1}\bigr) \tensor{I}
\\
&\qquad+\Bigl[
-
c_{0}^{2}
\order[*]{\rho}{1} \tensor{I}
+
\mu \Bigl(\nabla_{\bv{x}} \order[*]{\bv{v}}{1} + \nabla_{\bv{x}} \orderT[*]{\bv{v}}{1}\Bigr) + \mu_{\mathrm{b}} \bigl(\nabla_{\bv{x}} \scdot \order[*]{\bv{v}}{1}\bigr) \tensor{I}\Bigr]
\bigl[
(\nabla_{\bv{x}}\scdot\bv{\xi}) \tensor{I} - \transpose{\nabla_{\bv{x}}\bv{\xi}}
\bigr].
\end{aligned}
\end{equation}
We now observe that the term $\partial_{t}\order[*]{\bv{v}}{2}$ in Eq.~\eqref{eq: momentum balance second order} is of order higher than two.  In fact, using Eq.~\eqref{eq: v orders 2 def}, we have
\begin{equation}
\label{eq: order of dt v second order}
\partial_{t}\order[*]{\bv{v}}{2} = \nabla_{\bv{x}} \bv{v}_{\bv{\xi}} \partial_{t} \order{\bv{u}}{0}
+ \tensor{F}_{\bv{\xi}} \Bigl[\partial_{tt} \order{\bv{u}}{0} - \bigl(\nabla_{\bv{X}} \partial_{t}\order{\bv{u}}{0}\bigr) \bigl(\tensor{I} + \nabla_{\bv{X}} \order{\bv{u}}{0}\bigr)^{-1} \partial_{t} \order{\bv{u}}{0}\Bigr].
\end{equation}
As $\nabla_{\bv{x}} \bv{v}_{\bv{\xi}} = \mathcal{O}(\epsilon)$, $\partial_{t} \order{\bv{u}}{0} = \mathcal{O}(\epsilon^{2})$, and $\partial_{tt} \order{\bv{u}}{0} = \mathcal{O}(\epsilon^{4})$,
the above result implies that $\partial_{t}\order[*]{\bv{v}}{2} = \mathcal{O}(\epsilon^{3})$.  The latter result, combined with Eq.~\eqref{eq: BMass Lagrangian solution order 2 rate}, allows us to rewrite Eqs.~\eqref{eq: mass balance second order}--\eqref{eq: P second order} as follows:
\begin{equation}
\label{eq: mass and momentum balance second order simple}
\nabla_{\bv{x}}\scdot\order[*]{\bv{v}}{2} = 0
\quad\text{and}\quad
\nabla_{\bv{x}}\scdot \order[*]{\tensor{P}}{2} = \bv{0}.
\end{equation}
The first of Eqs.~\eqref{eq: mass and momentum balance second order simple} implies that $\order[*]{\bv{v}}{2}$ must be divergence-free.  In turn, this means that $\order[*]{\tensor{P}}{2}$ is determined by Eq.~\eqref{eq: P second order} only up to a hydrostatic contribution needed to accommodate the kinematic constraint in the first of Eqs.~\eqref{eq: mass and momentum balance second order simple} \citep{GurtinFried_2010_The-Mechanics_0}.  Therefore, recalling that $\order[*]{\rho}{0} = \rho_{0}$ and Eq.~\eqref{eq: BMass Lagrangian solution order 1}, the expression for $\order[*]{\tensor{P}}{2}$ that we will use in the formulation of the second-order problem will be as follows:
\begin{equation}
\label{eq: P second order simple}
\begin{aligned}[b]
\order[*]{\tensor{P}}{2} &= 
- q \tensor{I}
-\tfrac{1}{2}
c_{0}^{2}
\rho_{0} 
\Bigl[
(\nabla_{\bv{x}}\scdot\bv{\xi})^{2}
+
\transpose{\nabla_{\bv{x}}\bv{\xi}}\colondot\nabla_{\bv{x}}\bv{\xi}
\Bigr]
\tensor{I}
+
\mu \Bigl(\nabla_{\bv{x}} \order[*]{\bv{v}}{2} + \nabla_{\bv{x}} \orderT[*]{\bv{v}}{2}\Bigr)
\\
&\qquad
-\mu \Bigl(\nabla_{\bv{x}} \order[*]{\bv{v}}{1} \nabla_{\bv{x}}\bv{\xi} + \transpose{\nabla_{\bv{x}}\bv{\xi}} \nabla_{\bv{x}} \orderT[*]{\bv{v}}{1}\Bigr)
-
\mu_{\mathrm{b}} \bigl( \transpose{\nabla_{\bv{x}}\bv{\xi}} \colondot \nabla_{\bv{x}}\order[*]{\bv{v}}{1}\bigr) \tensor{I}
\\
&\qquad
+\Bigl[
c_{0}^{2}
\rho_{0} (\nabla_{\bv{x}}\scdot\bv{\xi}) \tensor{I}
+
\mu \Bigl(\nabla_{\bv{x}} \order[*]{\bv{v}}{1} + \nabla_{\bv{x}} \orderT[*]{\bv{v}}{1}\Bigr) + \mu_{\mathrm{b}} \bigl(\nabla_{\bv{x}} \scdot \order[*]{\bv{v}}{1}\bigr) \tensor{I}\Bigr]
\bigl[
(\nabla_{\bv{x}}\scdot\bv{\xi}) \tensor{I} - \transpose{\nabla_{\bv{x}}\bv{\xi}}
\bigr],
\end{aligned}
\end{equation}
where $q(\bv{x},t)$ is a scalar Lagrange multiplier that is determined by enforcing the constraint in the first of Eqs.~\eqref{eq: mass and momentum balance second order simple}.  We should remark that, from a strictly analytical viewpoint, the above relationship is equivalent to
\begin{equation}
\label{eq: P second order simple discussion}
\begin{aligned}[b]
\order[*]{\tensor{P}}{2} &= 
- \hat{q} \tensor{I}
+
\mu \Bigl(\nabla_{\bv{x}} \order[*]{\bv{v}}{2} + \nabla_{\bv{x}} \orderT[*]{\bv{v}}{2}\Bigr)
\\
&\qquad
-
\mu \Bigl(\nabla_{\bv{x}} \order[*]{\bv{v}}{1} \nabla_{\bv{x}}\bv{\xi} + \transpose{\nabla_{\bv{x}}\bv{\xi}} \nabla_{\bv{x}} \orderT[*]{\bv{v}}{1}\Bigr)
+
\mu (\nabla_{\bv{x}}\scdot\bv{\xi}) \Bigl(\nabla_{\bv{x}} \order[*]{\bv{v}}{1} +  \nabla_{\bv{x}} \orderT[*]{\bv{v}}{1}\Bigr)
\\
&\qquad
-\Bigl[
c_{0}^{2}
\rho_{0} (\nabla_{\bv{x}}\scdot\bv{\xi}) \tensor{I}
+
\mu \Bigl(\nabla_{\bv{x}} \order[*]{\bv{v}}{1} + \nabla_{\bv{x}} \orderT[*]{\bv{v}}{1}\Bigr) + \mu_{\mathrm{b}} \bigl(\nabla_{\bv{x}} \scdot \order[*]{\bv{v}}{1}\bigr) \tensor{I}\Bigr]
\transpose{\nabla_{\bv{x}}\bv{\xi}}.
\end{aligned}
\end{equation}
where $\hat{q}$ is the sum total of $q$ and all other hydrostatic contributions.  However, depending on the numerical method used to solve the problem, and on the corresponding discretization strategy, the two expressions are not equivalent from a numerical viewpoint.  As discussed later, our numerical scheme is a mixed \ac{FEM} formulation implemented via traditional Lagrange polynomials for the velocity field and the Lagrange multiplier $q$.  For said numerical approach, we have found that Eq.~\eqref{eq: P second order simple} leads to results without spurious numerical oscillations.

We now proceed to time average the equations developed so far.  The absence of time derivative in the governing equations of the second-order problem makes the time averaging operation straighforward and transparent.  In particular, we recall that the time average operation commutes with the spatial differentiation operators.  Furthermore, invoking Eq.~\eqref{eq: LLM v v*2 rel}, we recall that the time averaging of $\order[*]{\bv{v}}{2}$ yields the Lagrangian mean velocity.  Finally, as argued in Section~\ref{sec: time averaging and more} on p.~\pageref{sec: time averaging and more}, we recall that the time averaging operation is equivalent to a change of temporal scale from the fast to the slow time scale.  Hence, upon time averaging, the time-averaged second-order problem is governed by the following equations:
\begin{equation}
\label{eq: mass and momentum balance second order simple time averaged}
\nabla_{\bv{x}}\scdot \LM{\bv{v}} = 0
\quad\text{and}\quad
\nabla_{\bv{x}}\scdot \langle\order[*]{\tensor{P}}{2}\rangle = \bv{0},
\end{equation}
with
\begin{equation}
\label{eq: P second order simple time averaged}
\begin{aligned}[b]
\langle\order[*]{\tensor{P}}{2}\rangle &=
-\langle q\rangle \tensor{I}
+ \mu \Bigl[\nabla_{\bv{x}} \LM{\bv{v}} + \transpose{\bigl(\nabla_{\bv{x}}\LM{\bv{v}}\bigr)}\Bigr]
\\
&\quad
+\tfrac{1}{2}
c_{0}^{2}
\rho_{0} 
\Bigl\langle
(\nabla_{\bv{x}}\scdot\bv{\xi})^{2}
-
\transpose{\nabla_{\bv{x}}\bv{\xi}}\colondot\nabla_{\bv{x}}\bv{\xi}
\Bigr\rangle
\tensor{I}
+ \mu_{\mathrm{b}}
\Bigl\langle
\nabla_{\bv{x}} \scdot \bv{\xi}
\nabla_{\bv{x}} \scdot \order[*]{\bv{v}}{1}
-\transpose{\nabla_{\bv{x}}\bv{\xi}} \colondot \nabla_{\bv{x}}\order[*]{\bv{v}}{1}
\Bigr\rangle \tensor{I}
\\
&\quad
+\mu
\Bigl\langle
\nabla_{\bv{x}}\scdot\bv{\xi}\Bigl(\nabla_{\bv{x}} \order[*]{\bv{v}}{1} + \nabla_{\bv{x}} \orderT[*]{\bv{v}}{1}\Bigr) 
- \nabla_{\bv{x}} \order[*]{\bv{v}}{1} \nabla_{\bv{x}}\bv{\xi}
- \transpose{\nabla_{\bv{x}}\bv{\xi}} \nabla_{\bv{x}} \orderT[*]{\bv{v}}{1}
\Bigr\rangle
\\
&\quad
-
\Bigl\langle
\Bigl[
c_{0}^{2}
\rho_{0} (\nabla_{\bv{x}}\scdot\bv{\xi}) \tensor{I}
+
\mu 
\Bigl(\nabla_{\bv{x}} \order[*]{\bv{v}}{1} + \nabla_{\bv{x}} \orderT[*]{\bv{v}}{1}\Bigr) + \mu_{\mathrm{b}} \bigl(\nabla_{\bv{x}} \scdot \order[*]{\bv{v}}{1}\bigr) \tensor{I}\Bigr]
\transpose{\nabla_{\bv{x}}\bv{\xi}}
\Bigr\rangle,
\end{aligned}
\end{equation}
where $\order[*]{\bv{v}}{1} = \partial_{t} \bv{\xi}$ and therefore the last three lines of the above expressions are reduced to a single second-order tensor known from the first-order solution.  As far as the boundary conditions are concerned, these have already been stated in Eq.~\eqref{eq: vL boundary condition}.  The main observation concerning the present formulation is that, granted the kinematic framework we have laid out, the boundary conditions for the second-order problem are \emph{exact} as opposed to approximate as they appear in the purely Eulerian formulations of the second-order problem \citep[see, e.g.,][]{koster2007numerical}.

The formulation of the second-order problem in the proposed \ac{ALE} framework has yielded a (slow) time-independent problem at the second-order.  If there are no bodies immersed in the fluid and convected by it, then the second-order problem is intrinsically \emph{steady}.  This is in contrast to the Eulerian formulations found in the literature \citep[see, e.g.,][]{muller2012numerical,koster2007numerical}, which first derive a time-dependent second-order problem even upon time averaging and then consider an associated steady problem by setting the time derivatives of the primary unknowns to zero.  It should also be remarked that our result is  consistent with the analysis by~\citet{Xie2014multiscale}, which, in a regime like the one considered here, also derives an intrinsically time-independent second-order problem. This is an important observation in light of the recent studies which consider a time-dependent second-order case using Eulerian formulation~\citep{muller2015theoretical}. We believe the presence of the time-dependent terms at the second-order in Eulerian formulation to be suspicious in that it is derived by a time averaging operation that is somewhat ambiguous as to how it transitions from a time scale to another.  When the time-scale separation is explicitly defined, as was done here and in~\citet{Xie2014multiscale}, the analysis of the motion leads to the conclusion that the second-order problem is not time-dependent at the slow time scale.  This is also consistent with the fact that, whether in our formulation or in purely Eulerian ones, the forcing terms appearing in the equations, along with the boundary conditions, are time independent.

If an object is immersed in the fluid, the corresponding streaming flow, in general, would not be steady.  Provided that a more careful analysis of this case is needed, the present formulation suggests that one can proceed to update the position of the immersed object on the slow time by integrating the Lagrangian mean velocity over time intervals smaller than $\Pi/\epsilon^{2}$.  In the results section of the paper, we have included an example of the tracking of the displacement of a portion of the fluid using the immersed method discussed in \citet{Heltai2012Variational-Imp0}.  Applications with immersed solid objects in a streaming flow are currently being developed and will be presented in future publications.

As a final comment on the the governing equations of the second-order problem, we note that the first of Eq.~\eqref{eq: mass and momentum balance second order simple time averaged} is consistent with Eq.~\eqref{eq: volume preserving slow motion}, the latter pertaining to the zeroth-order problem.  That is, if we view the Lagrangian mean velocity as the effective velocity field of the motion at the slow time scale, requiring that the velocity field in question is divergence free is an expression of the fact that the slow motion is volume preserving as was argued after Eq.~\eqref{eq: volume preserving slow motion}.

\section{Numerical scheme implementation}
\label{sec: NumScheme}
Within the given framework, the solution of the zeroth-order problem is trivial and it has already been obtained.

As far as the first-order problem is concerned, this is intended to identify the system's response to a harmonic excitation.  Hence, as is typical in these cases, we seek a time-harmonic solution of the form
\begin{equation}
\label{eq: time harmonic solution form}
\bv{\xi}(\bv{x},t) = \hat{\bv{\xi}}(\bv{x}) e^{\ii \omega t},
\end{equation}
where $\hat{\bv{\xi}}(\bv{x})$ is a time independent complex-valued vector field of space only. Substituting this expression into the problem's governing equations, and as already discussed in Section~\ref{subsec: acoustic subproblem} on p.~\pageref{subsec: acoustic subproblem}, we obtain a problem that is formally similar to a traditional isotropic static linear elasticity problem with complex coefficients.  As such, it can be solved with a vast array of methods.  We have solved the first-order problem using a traditional \ac{FEM}, as demonstrated in \citet{Hughes2000The-Finite-Element-0}.  Granted that the solvability of the problem is dictated by the value of $\omega$ in relation to the physical parameters $\rho_{0}$, $\mu$, and $\mu_{b}$, for those values of $\omega$ for which a solution does exists, this solution can be very effectively approximated within traditional continuous piece-wise Lagrange polynomial finite element spaces \citep[cf.][]{Hughes2000The-Finite-Element-0}.  From a numerical viewpoint, the problem we obtain in our formulation is very different, and somewhat simpler, than what is typically obtained in Eulerian formulations such as that in \citet{koster2007numerical}.  In the referenced work, the problem has the mathematical structure of a Stokes problem, and therefore with two unknown fields (namely pressure and velocity). The numerical solution of such a problem is typically tackled via a mixed \ac{FEM} scheme employing structured elements such as the Taylor-Hood element \citep{koster2007numerical}.  In our case, the problem is stated in terms of a single field, namely $\bv{\xi}$, with consequent simpler economy in terms of total number of degrees of freedom to achieve a given accuracy, and with the possibility of choosing simpler and more easily found finite elements.

As far as the second-order problem is concerned, because this is time-independent, the problem reduces to a traditional steady Stokes flow problem \citep[cf.][]{Hughes2000The-Finite-Element-0}.  The solvability of this problem is well established \citep[see, e.g.,][]{Hughes2000The-Finite-Element-0,koster2007numerical,Brezzi1991Mixed-and-Hybrid-0} and we defer to the cited sources for formal existence and stability results.  We have adopted a standard approach of using a composite element with proven stability properties, such as, for 2D problems, $\mathcal{P}1$-$\mathcal{P}2$ or $\mathcal{P}2$-$\mathcal{P}3$, these being triangular elements with Lagrange polynomials for the pressure and the velocity fields of order $1$ and $2$, respectively, or order $2$ and $3$ respectively.

We do present the solution of a case in which the displacement of a sub-body of fluid is determined after the second-order problem is solved.  This consists of integrating Eq.~\eqref{eq: LLM v  u0 rel order of magnitude} (neglecting terms of order higher than $2$).  We note that the solution of this problem is predicated on a time integration at the slow time-scale in which, at each time step, one must sequentially solve both first- and second-order problems.  To achieve this, we have implemented the immersed \ac{FEM} discussed in \citet{Heltai2012Variational-Imp0}, to which we defer for formal details.  We remark that, to the best of the authors' knowledge, this is a completely new result in microacoustofluidics, one that opens the door to the fluid-structure interaction study of solid deformable objects immersed in a streaming flow. 

All the numerical solutions presented in the results section of this article were obtained for two-dimensional problems via the commercial finite element software \citet{COMSOL52}. We note that no commercial modules exist in \comsol for the solution of the problems formulated in this paper.  We have used \comsol as a high-level programming environment to create our own implementation of the problems in question.  Specifically, we have repeatedly used the  \emph{Weak PDE} interface available in \comsol to implement custom \ac{FEM}s.  As for the time integration for the reconstruction of the zeroth-order displacement field (the problem solved via the immersed \ac{FEM}), we have used the default framework available in \comsol consistsing of the method of lines \citep{Schiesser1991The-Numerical-Method-0} with the nonlinear solver in IDAS \citep{Hindmarsh2005SUNDIALS-Suite-0}.  This solver offers variable-order, variable-step-size \ac{BDF} for the solution of \ac{DAE} systems of the type $F(t,y,y',p) = 0$. The \ac{BDF} method was configured so as to allow orders $1$--$5$.

\section{Numerical Results}
\label{sec: results}
In this section, we present numerical results obtained with our formulation for different actuation functions and devices of varying size with respect to the wavelength of the acoustic actuation employed. We compare our results with those obtained via two other common approaches discussed in the literature. Specifically, we consider results from Eulerian formulations with two distinct approaches to the enforcement of boundary conditions.  To help in the presentation of this comparative analysis, we will resort to labels by which we identify the methodology employed in obtaining the results.  The labels in question are as follows:
\begin{enumerate}
\item
We will use the label `\ac{ALE}' to identify results obtained with our approach. We recall that for our formulation the principal unknown of the second-order problem is the Lagrangian mean particle velocity and that, as argued earlier in the paper, the boundary value of this quantity is equal to zero under no slip conditions.

\item
We will use the label `$\EMC$' to denote an Eulerian formulation in which the velocity boundary conditions for the second-order problem are those proposed in \citet{Bradley1996Acoustic-Stream0,bradley2012acoustic}.  This is the formulation analyzed formally and numerically in \citet{koster2007numerical}, and for which the boundary (in the second-order problem) is viewed as \emph{moving} according to a second-order expansion of the boundary displacement for the physical problem.  The label's subscript stands for `moving boundary conditions'.

\item 
We will use the label `\EZBC' to denote an Eulerian formulation in which the velocity boundary conditions for the second-order problem are simply taken as homogeneous and of Dirichlet type, i.e., the boundary value of the velocity field is set to zero. The label's subscript stands for `zero boundary conditions'. While this formulation has been used by various authors, we will specifically refer to the work in \citet{muller2012numerical}.
\end{enumerate}

\subsection{Rectilinear actuation}
\label{subsec: BAW}

\begin{figure}
\centering
\includegraphics[scale=0.35]{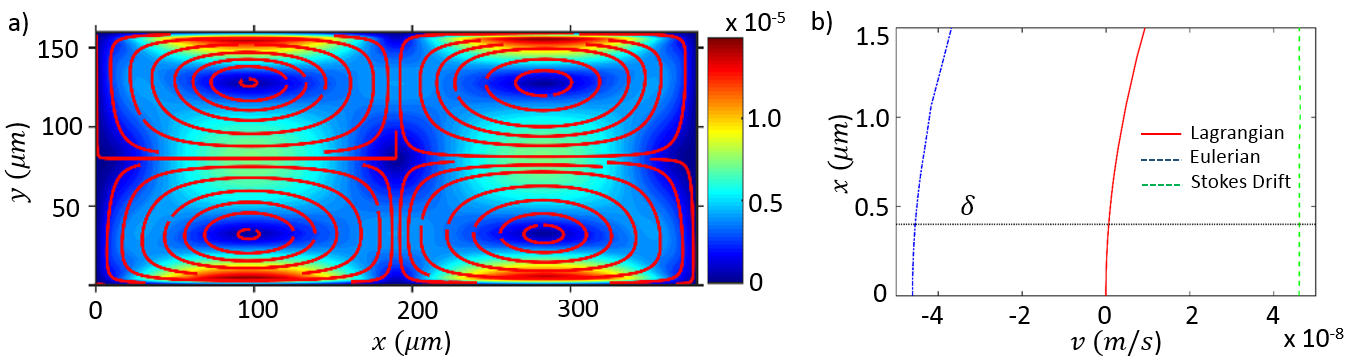}
\caption{(a) \ac{ALE} formulation: Density plot of the magnitude of the Lagrangian mean velocity for a bulk acoustic wave device, where the vertical walls are actuated as indicated in Eqs.~\eqref{eq: rectilinear actuation 1}--\eqref{eq: rectilinear actuation 4}. The red lines represent the streamlines of the Lagrangian mean velocity. (b) \ac{ALE} formulation: Plot of the $x$ component of different velocity fields away from the left wall (at $x=0$) along a horizontal line at y=$100~\microns$. At the oscillating wall (i.e. $x=0$), the Eulerian velocity is equal and opposite to the Stokes drift, thereby yielding a zero Lagrangian velocity at the oscillating wall representing no flow across the wall.}
\label{fig: Rectilinear}
\end{figure}

This numerical test is motivated by the \ac{BAW} devices considered by~\citet{muller2012numerical}.
With reference to Fig.~\ref{fig: Rectilinear}(a),
the computational domain consists of a rectangular 2D fluid-filled cavity with length $L = \np[\mu m]{380}$ and height $H=\np[\mu m]{160}$, where the left and the right walls are actuated harmonically along the $x$-direction. The actuation frequency considered is $f = \np[MHz]{1.97}$ which corresponds to a half-wave resonance mode for the channel dimensions considered. Using a Cartesian coordinate system with origin at the lower left corner of the domain, the actuation displacement at the oscillating walls is given by
\begin{alignat}{2}
\label{eq: rectilinear actuation 1}
u_{x}(\bv{x},t) &= d_{0} e^{\ii \omega t}\quad&
&\text{for $x = 0,L$ and $0 \leq y \leq H$},
\\
\label{eq: rectilinear actuation 2}
u_{y}(\bv{x},t) &= 0\quad&
&\text{for $x = 0,L$ and $0 \leq y \leq H$},
\\
\label{eq: rectilinear actuation 3}
u_{x}(\bv{x},t) &= 0\quad&
&\text{for $y = 0,H$ and $0 \leq x \leq L$},
\\
\label{eq: rectilinear actuation 4}
u_{y}(\bv{x},t) &= 0\quad&
&\text{for $y = 0,H$ and $0 \leq x \leq L$},
\end{alignat}
where $u_{x}$ and $u_{y}$ are the displacement components in the $x$ and $y$ directions, respectively, $d_0 = \np[m]{1e-10}$ is the actuation amplitude, and $\omega=2\pi f$ is the angular frequency. The fluid properties are taken to be those of water and are same as those considered by~\citet{muller2012numerical}. 

The results in Fig.~\ref{fig: Rectilinear} were obtained with the \ac{ALE} method. Fig.~\ref{fig: Rectilinear}(a) shows a density plot of the magnitude of Lagrangian mean velocity $\LM{\bv{v}}$.  The lines superimposed to the density plot trace the streamlines of $\LM{\bv{v}}$. Our results indicate the presence of four vortices inside the channel, similar to those reported by~\citet{muller2012numerical} via a $\EZBC$ formulation. In fact our results are very similar to those in \citet{muller2012numerical} in the bulk of the channel.  However, they differ significantly near the oscillating walls.  Before illustrating the differences, it is important to keep in mind that differences and similarities of results need to be understood \emph{in context}, i.e., with an awareness of the size of the device and the forcing frequency and amplitude.  With this in mind, it is generally accepted that  a second-order Eulerian velocity field does not accurately represent particles trajectories.  Accurate predictions of these are obtained by summing to the Eulerian velocity field the Stokes drift \citep{Buhler2009Waves-and-Mean-0}.  The fact that for the case under discussion our results match those in \citet{muller2012numerical}, the latter obtained without Stokes drift correction, indicates that in this particular problem the correction in question may be negligible in the bulk of the channel.  To verify this conjecture we show in 
Fig.~\ref{fig: Rectilinear}(b) the plot of the $x$ component of Lagrangian mean velocity, the Eulerian velocity, and the Stokes drift near the oscillating left wall. Again, the plots shown in this figure were obtained via the \ac{ALE} formulation.  It can be observed that the Eulerian velocity and Stokes drift are equal in magnitude but opposite in direction at the oscillating wall, thereby giving a zero Lagrangian velocity at the oscillating wall. We also note that while the Stokes drift is equal and opposite to the Eulerian velocity field at the oscillating boundaries, the Stokes drift decay significantly away from the oscillating walls and the maximum value of Stokes drift is observed to be one order of magnitude smaller than the maximum value of Eulerian velocity field in the bulk. Therefore, the contribution of Stokes drift to the net Lagrangian field is indeed negligible in the bulk of the channel, resulting in no significant differences between the Eulerian and the Lagrangian velocity fields in the bulk of the channel. Thus the results obtained from the \ac{ALE} formulation do not show any significant differences in the bulk of the channel compared to those reported by~\citet{muller2012numerical} for this specific case.

What is remarkable about this conclusion is that, in reality, the boundary conditions in the $\EZBC$ formulation are known to pose difficulties in their physical meaning.  In fact, owing to the rectilinear nature of the actuation displacement profile, there is no slip velocity along the oscillating walls and only a non-zero transverse component of Stokes drift exist at the oscillating walls. From a physical perspective, a zero Eulerian velocity at the oscillating walls along with a nonzero Stokes drift correction leads to the conclusion that fluid mass is being convected through the oscillating walls. By contrast, this contradiction is completely avoided both in the current \ac{ALE} formulation and in the $\EMC$ formulation.  In the latter, the boundary value of the Eulerian velocity is not zero, but it is equal and opposite to the Stokes drift thereby producing a zero Lagrangian mean velocity.  It is important to note that, from a numerical viewpoint, the \ac{ALE} and the $\EMC$ formulations do not yield identical results.  This aspect will be considered in greater detail in Section~\ref{subsection: numerical performance} on page~\pageref{subsection: numerical performance}.


\subsection{Non-Rectilinear actuation}
In the previous section we mentioned the fact that, in some cases, the Eulerian velocity can be taken as an accurate enough descriptor of the mean particle velocity.  In this section we present a more in-depth analysis of this question and we will point out that there are a number of applications of technological relevance in which the Eulerian velocity is not a good enough descriptor of particle motion especially as the size of the device becomes of the same order of magnitude or smaller than the actuation wave length.

\subsubsection{Case I: $L \gg \lambda$}
\label{sec: Lmuchlesslamnda}

\begin{figure}
\centering
\includegraphics[scale=0.4]{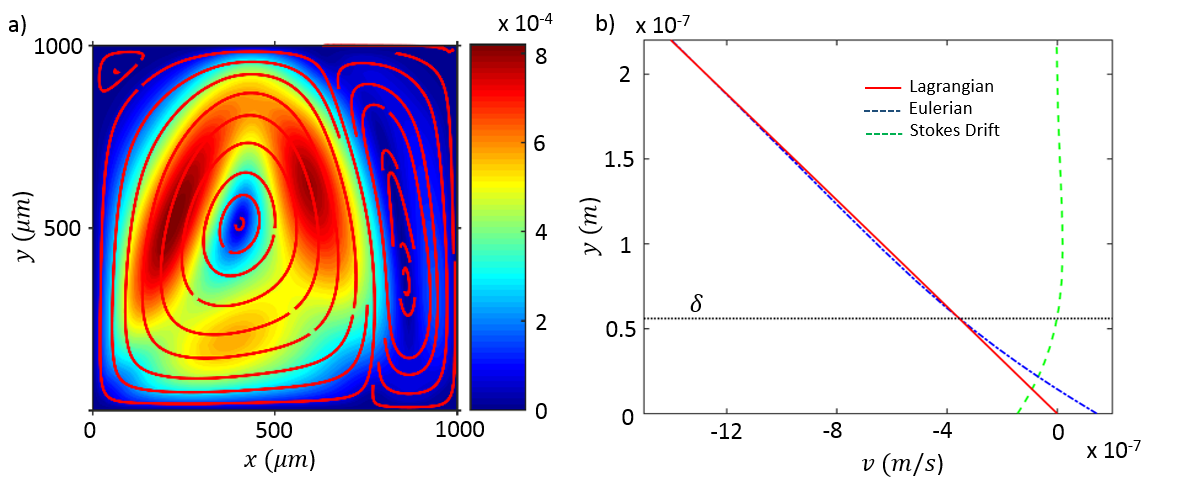}
\caption{(a) Density plot of the magnitude of the Lagrangian mean velocity $\LM{\bv{v}}$ for a device that is large compared to the acoustic wavelength.  The device is actuated only through the bottom wall. The red lines represent the streamlines of $\LM{\bv{v}}$. (b) Plot of the $x$-component of the $\LM{\bv{v}}$ along with the $x$ component of the Stokes drift and the Eulerian velocity as a function of $y$ in a region close to the bottom wall on the order of the boundary layer thickness $\delta = 5.6\times10^{-8} m$. away from the oscillating wall along a vertical line at $x = \np[\mu m]{400}$. At the oscillating wall (i.e. $y=0$), the Eulerian velocity is equal and opposite to the Stokes drift, thereby yielding a zero Lagrangian velocity at the oscillating wall representing no flow across the wall.}
\label{fig: Koster SAW}
\end{figure}

Here, we consider a traveling \ac{SAW} device, similar to the one considered by~\citet{koster2007numerical}. Referring to Fig.~\ref{fig: Koster SAW}(a)
the computational domain consists of a 2D fluid-filled square with side of length equal to $\np[mm]{1}$.  The vertical and top sides of the square are stationary whereas the bottom side is subject to a displacement actuation as follows:
\begin{equation}
u_{x}(\bv{x}, t) = d_{0} e^{-C_{d} x} \sin(2 \pi x/\lambda) e^{\ii \omega t}
\quad\text{and}\quad
u_{y}(\bv{x},t) = -d_{0} e^{-C_{d} x} \cos(2 \pi x/\lambda) e^{\ii \omega t},
\end{equation}
where $u_{x}$ and $u_{y}$ are the displacement components along the $x$ and $y$ direction, respectively, $x \in [0,\np[mm]{1}]$, $y = 0$,  $C_{d} = \np[m^{-1}]{6080}$ is the decay coefficient, $d_{0}= \np[m]{1e-10}$ is the actuation amplitude, and $\omega=2\pi f$ is the angular frequency. The relevant material parameters for the numerical test are same as those considered by~\citet{koster2007numerical}, with the actuation frequency $f = \np[MHz]{100}$. For the commonly used Lithium-Niobate substrate in these devices, this frequency corresponds to a wavelength of $\lambda\approx \np[\mu m]{38}$, which is much smaller than the device dimensions considered. We note that the actuation function considered here differs from that considered in the previous section in that it is not uniform along the oscillating wall. Hence, unlike the previous case, it results in a non-zero lateral component of the Stokes drift along the oscillating boundary.  In $\EMC$ formulations this fact is accounted for by including said lateral component of velocity in the second-order problem boundary conditions. Fig.~\ref{fig: Koster SAW}(a) shows the density plot of $\|\LM{\bv{v}}\|$ and the  lines denote the streamlines of $\LM{\bv{v}}$ obtained from the ALE formulation.  Along with the calculation using our \ac{ALE} formulation we repeated the analysis of this problem using both a $\EMC$ formulation \citep[as in][]{koster2007numerical} and a  $\EZBC$ formulation \citep[as in][]{muller2012numerical}. Similar to the case considered in the previous section, no significant differences were observed between the Lagrangian mean velocity field and the Eulerian velocity field in the bulk of the channel.  To explain this finding, in Fig.~\ref{fig: Koster SAW}(b) we plot the $x$-component of $\LM{\bv{v}}$, the Stokes drift, and the corresponding Eulerian velocity as determined by our \ac{ALE} formulation as a function of $y$ along a line at $x = \np[\mu m]{400}$ and in the proximity of the bottom wall.   The horizontal dashed line represents the edge of the boundary layer with thickness $\delta = 5.6\times10^{-8} m$.  What we find is that the the Lagrangian mean and Eulerian velocities differ significantly inside the boundary layer compared, but the difference becomes rapidly negligible outside the boundary layer. This is due to the fact that, in the bulk of the channel, the maximum value of the Stokes drift is about two orders of magnitude smaller than the maximum value of Eulerian and Lagrangian velocity. From a physical perspective, in devices that are much larger than the wavelength, such as the one considered here, the largest values of the (magnitude of the) Eulerian velocity field occur inside the bulk of the domain well outside the boundary layer so that the distinction between the Lagrangian mean velocity and the Eulerian velocity is not appreciable.   Next, we will see that this is no longer the case for 
devices whose characteristic size is comparable to the actuation wavelength.

\subsubsection{Case II: $L \approx \lambda$}
\label{subsec: Lcomplambda}

\begin{figure}
\centering
\includegraphics[scale=0.45]{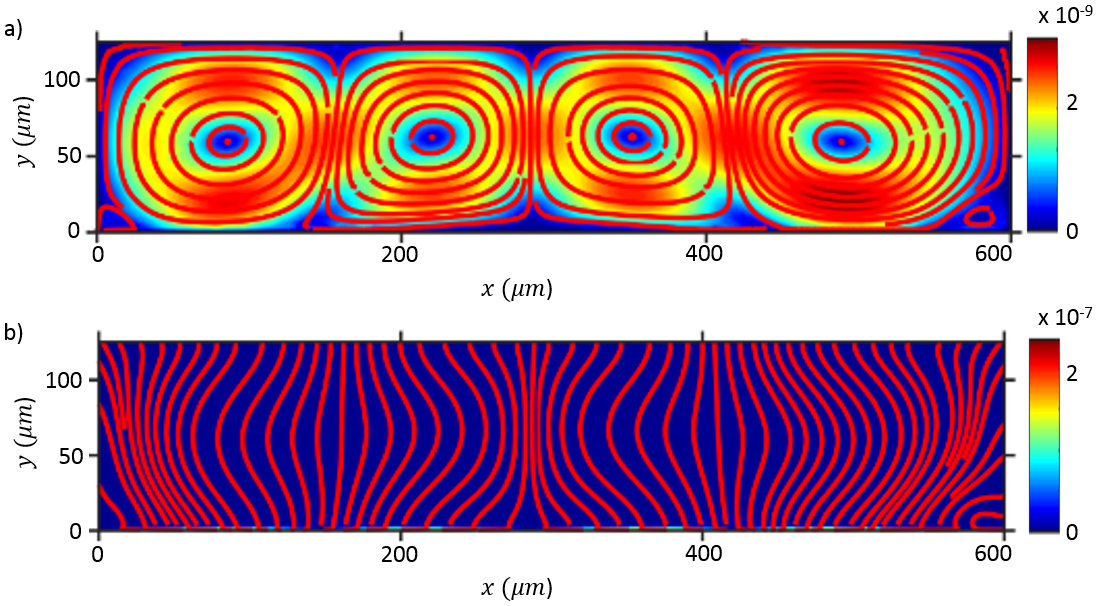}
\caption{\ac{ALE} formulation: Density plots of the magnitude of the Lagrangian mean velocity (a) and of the Eulerian velocity (b) for a device with dimensions comparable to the acoustic wavelength.  The bottom wall is actuated. The lines represent the streamlines of the respective velocities. The Lagrangian velocity streamlines do not cross the channel walls indicating no mass flow across the channel walls, while the Eulerian velocity streamlines cross the channel walls as though there were an outflow across the channel walls.}
\label{fig: L approx ambda case}
\end{figure}

We now consider a \ac{SAW} device where the channel dimensions are comparable to the wavelength of the acoustic actuation employed, similar to the one considered by~\citet{nama2015numerical}. With reference to Fig.~\ref{fig: L approx ambda case}
the computational domain consists of a 2D fluid-filled rectangular cavity with length $L = \np[\mu m]{600}$ and height $H = \np[\mu m]{125}$, where the bottom wall is actuated by a \ac{SAW}. The actuation frequency considered is $f = \np[MHz]{6.65}$ which corresponds to a wavelength of $\lambda = \np[\mu m]{600}$ for a Lithium-Niobate substrate. The actuation displacement profile considered is as follows:
\begin{equation}
u_{x}(\bv{x}, t) = 0
\quad\text{and}\quad
u_{y}(\bv{x},t) = d_{0} e^{-C_{d} x} \sin(2 \pi x/\lambda) e^{\ii \omega t},
\end{equation}
while the material parameters considered are same as those considered in the previous section, with the exception of the decay coefficient that here is $C_{d} = \np[m^{-1}]{116}$, corresponding to $f = \np[MHz]{6.65}$. The primary difference between the current case and the one considered in the previous section is the relative size of the device compared to the acoustic wavelength. Another difference is the fact that the walls of the microchannel consist of \ac{PDMS}  and have been modeled using the impedance boundary conditions similar to those employed by~\citet{nama2015numerical}. We also note that in terms of the actuation displacement, this case is also similar to the progressive surface wave radiator case considered~\citet{bradley2012acoustic}, albeit in a confined channel with dimensions comparable to the wavelength. As noted in the previous section, while this actuation function has only a non-zero transverse component, the non-uniformity of this function along the oscillating wall still results in a non-zero lateral component of Stokes drift along the oscillating boundary. 
Fig.~\ref{fig: L approx ambda case}(a) shows a density plot of the magnitude of the Lagrangian mean velocity $\LM{\bv{v}}$ along with the streamlines of $\LM{\bv{v}}$. Four streaming vortices are observed along the length of the channel, with the maximum value of $\|\LM{\bv{v}}\|$ being $\np[m/s]{3e-9}$. Fig.~\ref{fig: L approx ambda case}(b) shows the plot of the corresponding Eulerian velocity field where no vortices are observed and the maximum value of the norm of the Eulerian velocity is observed to be two orders of magnitude larger than that of $\|\LM{\bv{v}}\|$.

\begin{figure}
\centering
\includegraphics[scale=0.45]{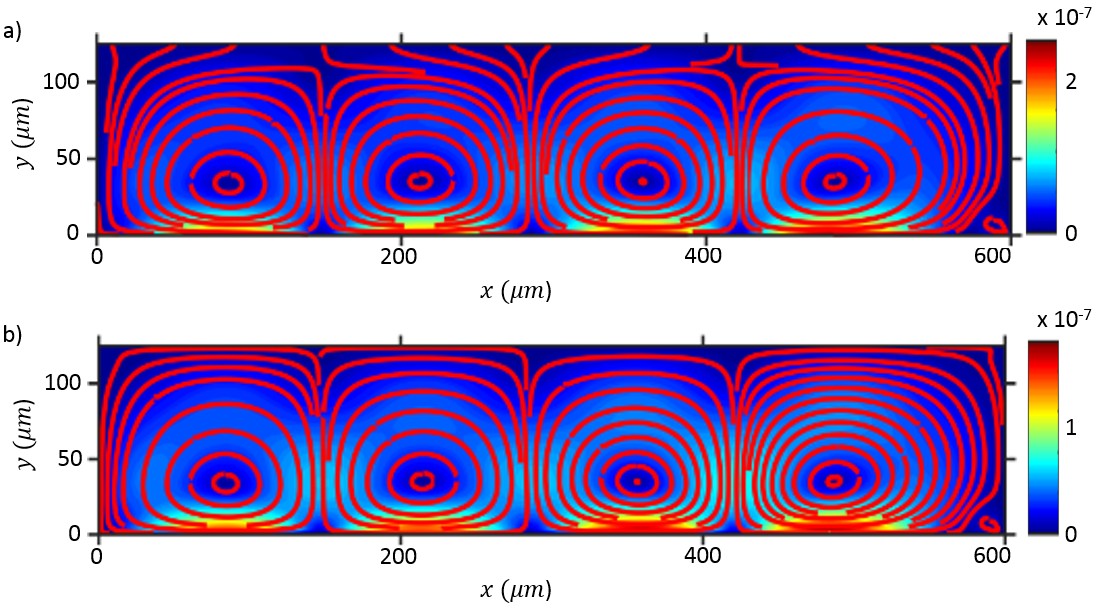}
\caption{$\EZBC$ formulation: Density plots of the magnitude of the (a) Lagrangian velocity and (b) Eulerian velocity for a device with dimensions comparable to the acoustic wavelength.  The bottom wall is actuated. These results are to be contrasted with those in Fig.~\ref{fig: L approx ambda case}.  The $\EZBC$ formulation (results in this figure) yields results that we believe are erroneous.  The lines represent the streamlines of the respective velocities. Here the Lagrangian velocity streamlines cross the channel walls indicating an unphysical mass flow across the channel walls. By contrast, consistent with the boundary conditions imposed in $\EZBC$ formulation, here the Eulerian velocity streamlines do not cross the channel.  These results are not supported by experimental evidence reported in \citet{Barnkob2016}.}
\label{fig: SAW_Muller}
\end{figure}

To investigate the effect of the boundary conditions for the second-order velocity, we also performed corresponding calculations for this case using a $\EZBC$ formulation. Figure~\ref{fig: SAW_Muller}(a) 
shows the density plot of the magnitude of the Lagrangian velocity obtained using an $\EZBC$ formulation where the lines denote the streamlines of said field. Similar to the results obtained with our \ac{ALE} formulation, four streaming vortices are observed along the length of the channel.  However, these are shifted closer to the oscillating wall when compared to those in our \ac{ALE} formulation. Furthermore, the maximum value of the norm of the Lagrangian velocity is approximately $\np[m/s]{2e-7}$.  Figure~\ref{fig: SAW_Muller}(b) shows the corresponding plot pertaining to the Eulerian velocity.  One immediate observation is that, in this case, the Lagrangian and Eulerian fields show no significant difference between them except at the boundaries, but both differ significantly from the corresponding results obtained using our \ac{ALE} formulation. Specifically, the Lagrangian flow field obtained from the $\EZBC$ formulation differs in three aspects:
\begin{enumerate}
\item
The Lagrangian flow field at the channel boundaries is non-zero, indicating an nonphysical mass flux across the oscillating boundary.

\item
The maximum of the norm of the Lagrangian velocity from the $\EZBC$ formulation is two orders of magnitude larger than the corresponding value from the \ac{ALE} formulation.

\item
The direction of the Lagrangian velocity from the $\EZBC$ formulation is opposite to that obtained via the \ac{ALE} formulation.
\end{enumerate}
Based on the observations from the preceding sections, the first discrepancy is expected since, in general, for actuation cases where the Stokes drift is non-zero at the oscillating walls, setting the Eulerian velocity field to zero at the oscillating wall automatically results in a non-zero Lagrangian mean velocity and hence a corresponding mass flux across the oscillating boundary. The last two discrepancies can be understood based on the observation that in this case the Stokes drift term significantly affects the flow field even outside the boundary layer and its strength is of same order of magnitude as that of the Eulerian flow velocity. The mutually opposite directions of the Eulerian velocity field and Stokes drift field causes them to effectively cancel each other, except inside the boundary layer, thereby yielding a much slower Lagrangian velocity field. Recent experimental results by \citet{Barnkob2016}, show that our \ac{ALE} formulation yields correct magnitude and direction of the Lagrangian velocity, resulting in particle trajectory predictions that are much closer to the experiments than other formulations both qualitatively and quantitatively.  A follow up paper with a detailed comparison between numerical predictions and experiments is in preparation.

\subsection{Remarks on the numerical implementation}
\label{subsection: numerical performance}
\begin{figure}
\centering
\includegraphics[scale=0.80]{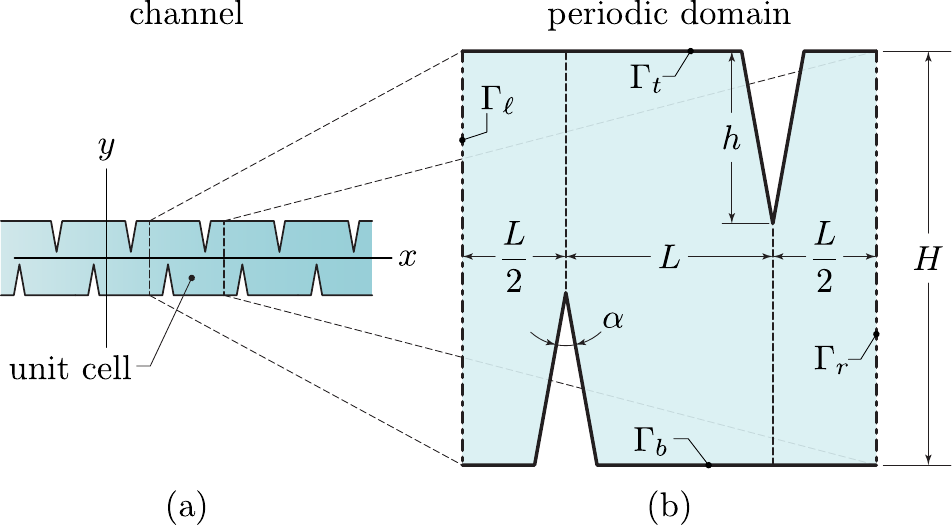}
\caption{Schematic design of a typical oscillating sharp-edge based device.}
\label{fig: Periodic_Cell}
\end{figure}

\begin{figure}
\centering
\includegraphics[scale=0.45]{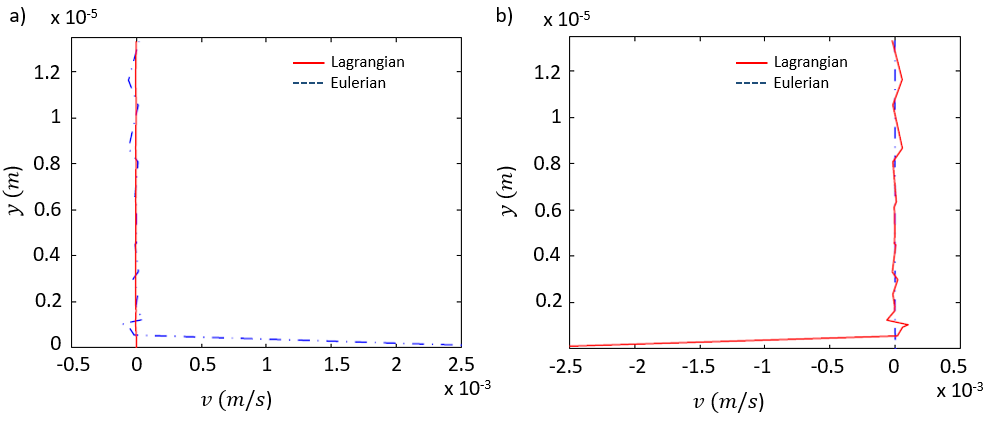}
\caption{Simulations pertaining to the sharp-edge device in Fig.~\ref{fig: Periodic_Cell}: $y$ component of the Lagrangian mean velocity of the and Eulerian velocity along a vertical line emanating from a sharp-edge tip obtained using (a) \ac{ALE} formulation and (b) $\EZBC$ formulation. }
\label{fig: SE_Comparison}
\end{figure}

\begin{figure}
\centering
\includegraphics[scale=0.4]{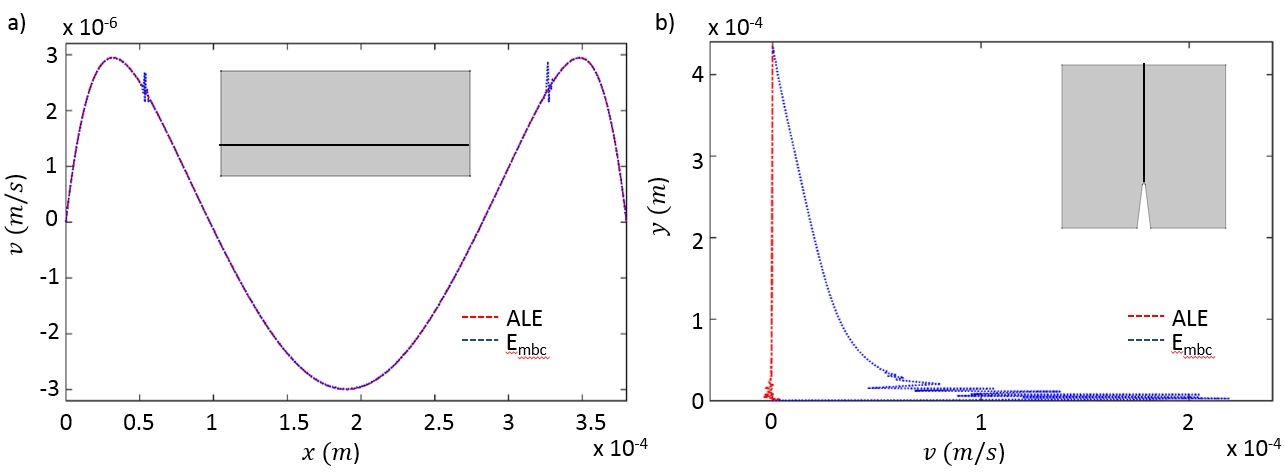}
\caption{(a) Plot of the $y$ component of the Lagrangian mean velocity obtained from the \ac{ALE} formulation and  the $\EMC$ formulation along a horizontal line (shown in the inset as a black line) at $y=H/3$ for the bulk acoustic wave device considered in Section~\ref{subsec: BAW} on page~\pageref{subsec: BAW}. (b) Plot of the $x$ component of the Lagrangian mean velocity obtained from the \ac{ALE} formulation and $\EMC$ formulation along a line parallel to the $y$ emanating from the sharp-edge tip towards the top wall as shown in the inset. As it computes $\LM{\bv{v}}$ directly, i.e., avoiding a postoprocessing involving the gradients of the first-order solution, the \ac{ALE} formulation yields smoother numerical results in both cases.}
\label{fig: NumericalPerformance}
\end{figure}

While the theoretical framework for our $\ac{ALE}$ formulation may appear more involved that the Eulerian setting of previous formulations, it is important to note that the boundary value problems actually solved both for the first- and the second-order problems are simpler that those of said formulations.  The solution of the first-order problem requires the solution of a linear problem in the displacement field $\bv{\xi}$ with no pressure or density constraints and the second-order problem has the structure of a steady Stokes problem.  In addition to these formal properties, the \ac{ALE} formulation provides the advantage of identifying the Lagrangian mean velocity field directly, i.e., without a postprocessing step that is susceptible to numerical aberration.  To illustrate this point, we consider the oscillating sharp-edge device considered by~\citet{nama2014investigation,nama2016investigation} and by~\citet{huang2013acoustofluidic,huang2015acoustofluidic,junahuang2014reliable,huang2015spatiotemporally}. Figure~\ref{fig: Periodic_Cell}(a) 
shows the schematic of a typical sharp-edge based acoustofluidic device consisting of protruded sharp-edges along the channel walls in a periodic fashion. Due to the inherent periodicity of the geometry as well as the actuation, we consider a periodic cell of the device as shown in Fig.~\ref{fig: Periodic_Cell}(b).  In our calculation, we have set $L = \np[\mu m]{600}$, $\alpha = 15^{\circ}$, $h = \np[\mu m]{200}$, and $H = \np[\mu m]{600}$. The top and the bottom channel walls (including the sharp-edges) denoted by $\Gamma_{t}$ and $\Gamma_{b}$, respectively, are assumed to undergo uniform harmonic motion along the $x$ direction with an actuation frequency $f = \np[kHz]{5}$. The side walls of the channel, denoted by $\Gamma_l$ and $\Gamma_r$, respectively, are subjected to periodic boundary conditions. To avoid the singularity associated with the second-order velocity field at the sharp-edge tip, we have rounded off the tip of the sharp-edges with a small radius of $\np[\mu m]{2}$. Figure~\ref{fig: SE_Comparison}(a) 
shows the plot of the $y$ component of the Lagrangian and the Eulerian velocities along a line parallel to $y$ axis moving away from the sharp-edge tip towards the top wall. Because the actuation consists of an oscillatory displacement purely parallel to the $x$ direction, the $y$ component of the particle velocity at the sharp edge is equal to zero.  The \ac{ALE} formulation, in which this boundary condition is applied directly, displays a mean Lagrangian velocity field consistent with the boundary condition in question.  However, what is more important, is that the component of $\LM{\bv{v}}$ obtained via the \ac{ALE} formulation is not characterized by numerical oscillations spanning individual elements in the simulation as is the case for the corresponding component of the Eulerian velocity.   Figure~\ref{fig: SE_Comparison}(b) displays the same result, this time obtained via the $\EZBC$ formulation.  As expected, here we obtained the reverse situation, that is, given a relatively smooth result for the Eulerian velocity, the reconstructed Lagrangian mean velocity suffers from severe numerical oscillations, often spanning individual elements.  The reason behind these oscillations in question is that the reconstruction of, say, the Lagrangian mean velocity from the Eulerian velocity requires the determination of Stokes drift, which, in turn contains spatial gradients of the first-order velocity solution, the latter, for harmonic solutions, being a close relative to $\nabla_{\bv{x}} \bv{\xi}$.  As the quality of the numerical results decreases (in general) with the order of differentiation, the choice of the approximation spaces for the solution then becomes crucial in achieving a desired accuracy in the post-processing operation at hand.  Consequently, if a simulation is meant to focus on particle velocities and trajectories, there is a definite advantage in choosing a formulation in which the Lagrangian mean velocity appears as one of the primary unknowns of the problem.  The reconstruction of the Lagrangian mean velocity from the Eulerian velocity field remains a delicate operation even when adopting the $\EMC$ formulation.  To illustrate this point, we present in Fig.~\ref{fig: NumericalPerformance}(a) 
a plot of the $y$ component of the Lagrangian mean velocity obtained from our \ac{ALE} formulation and with the $\EMC$ formulation for the simulation discussed in Section~\ref{subsec: BAW} on page~\pageref{subsec: BAW}.  Specifically, Fig.~\ref{fig: NumericalPerformance}(a) presents the plot of $\LM{v}_{y}$ along a horizontal line (shown in the inset as a black line) at $y = H/3$ (cf.\ Fig.~\ref{fig: Rectilinear}(a)).  In Fig.~\ref{fig: NumericalPerformance}(b) we report a result for the device shown in the inset but this time pertaining to the $x$ component of $\LM{\bv{v}}$ as obtained via our \ac{ALE} formulation and the $\EMC$ formulation.  In some sense, the case in Fig.~\ref{fig: NumericalPerformance}(b) is not surprising as the numerical aberration under discussion occurs in the proximity of a boundary with a re-entrant corner geometry.  That is, the reconstruction of $\LM{\bv{v}}$ from the Eulerian velocity as delivered by the $\EMC$ formulation is expected to be more susceptible to numerical aberration in the neighborhood of a high velocity gradient location.  What is more surprising is the result in Fig.~\ref{fig: NumericalPerformance}(a).  In this case the result coming from the $\EMC$ formulation is expected to be essentially identical to that coming from the \ac{ALE} formulation given that the Stokes drift is negligible in the bulk of the \ac{BAW} device under consideration.  So far we have outlined the more apparent numerical issues behind the difficulty of an accurate reconstruction of $\LM{\bv{v}}$ from the needed Eulerian fields.  In reality, at the root of these difficulties there is a more subtle analytical problem surrounding the determination of the inverse of the \ac{ALE} map $\bv{x} \mapsto \bv{x} + \bv{\xi}(\bv{x},t)$. From a theory of partial differential equations perspective, this issue is too technical to be addressed in this publication, but it has been mentioned by B\"uhler in the Lagrangian mean field theory section of \citet{Buhler2009Waves-and-Mean-0}. To clarify somewhat the issue at hand, one can consider the derivation of the Lagrangian mean velocity from within an Eulerian formulation, which we have reported in the Appendix in Eqs.~\eqref{eq: app vL classic 1}--\eqref{eq: velocityrelation}. With reference to Eq.~\eqref{eq: app vL classic 2} in particular, and consequently to Eq.~\eqref{eq: app vL classic 5}, we see that the concept of  Lagrangian mean velocity presupposes the existence of a smooth enough inverse of the aforementioned \ac{ALE} map.  It is not clear under what conditions one can guarantee the existence of this map for long enough time intervals.  With this in mind, and granted that this analytical problem is not answered by simply switching to an \ac{ALE} perspective as we have done, the approach proposed herein in which $\LM{\bv{v}}$ is a primary unknown (as opposed to a quantity computed in a post-processing step) does offer some advantages from a numerical viewpoint for simulations that focus on predicting particle trajectories, as is the case in the design of devices for the accurate placement of particles in the flow and for validation based on particle velocimetry experiments.

\subsection{Fluid Body Tracking}
\begin{figure}
\centering
\includegraphics[scale=0.50]{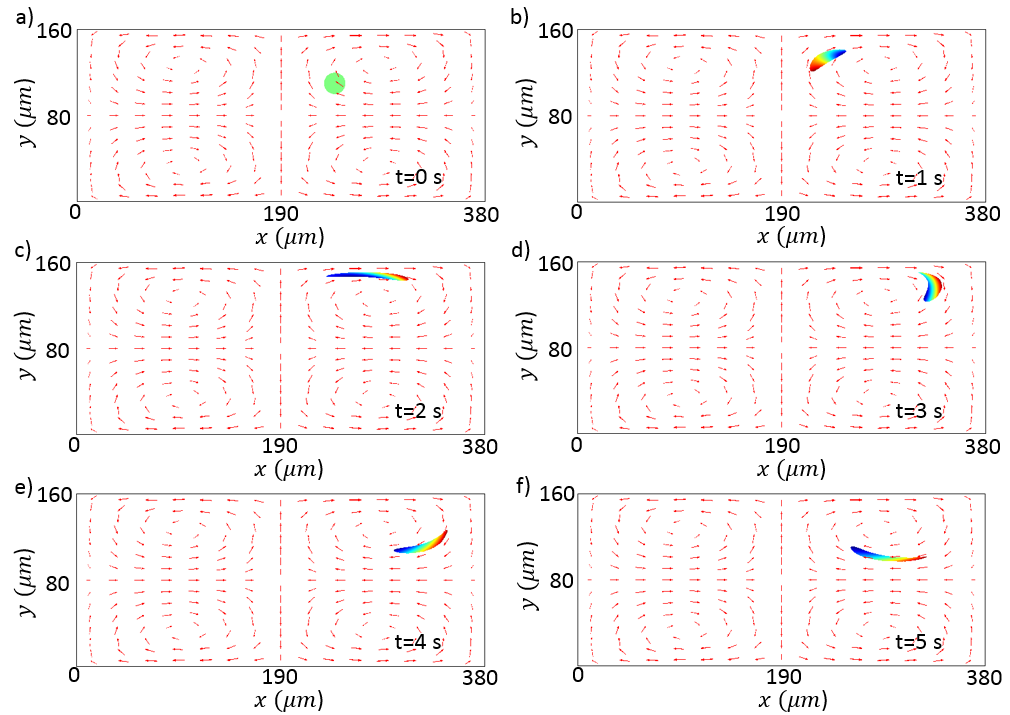}
\caption{Simulation showing the deformation of a continuous fluid sub-body originally occupying the circular region in frame (a).}
\label{fig: TimeLapse}
\end{figure}

The last set of numerical results we show concerns the tracking of a region of fluid within a streaming field.  This simulation is being reported here as a ``proof of concept'' in that it demonstrates one major application of the proposed \ac{ALE} formulation in the determination of the motion of solid bodies immersed in a streaming flow.  This is a problem that has been considered extensively in more traditional \ac{FSI} contexts \citep[see, e.g.,][]{Peskin_2002_The-immersed_0,Heltai2012Variational-Imp0,ZhangGay-2007-Immersed-finite-0}, but it has not been given a satisfactory treatment in microacoustofluidic applications.  The latter are of strong interest in acoustophoresis for controlling the motion and placement of immersed particles, whether soft or hard, and in applications for the manipulation of relatively large soft immersed objects such as cells.  Here we limit ourselves to illustrating the possibility of tracking the motion of an immersed body by tracking the motion of a connected fluid sub-body (an extensive analysis of the motion of soft immersed solid bodies by the authors is in preparation).  Figure~\ref{fig: TimeLapse}
shows several frames of a simulation in which we track the motion of the fluid sub-body originally occupying the circular domain shown in Fig.~\ref{fig: TimeLapse}(a).  The simulation was obtained by using the immersed method discussed in \citet{Heltai2012Variational-Imp0} for the integration of Eq.~\eqref{eq: LLM v  u0 rel order of magnitude} (including only terms up to order two).  The density plot over the tracked body displays the magnitude of the displacement.  The simulation in question is akin to determining the trajectory of a passive tracer where the tracer in question is not a point but a continuous region of fluid.  If the subbody in question consisted of, say, a soft elastic object, the computed displacement $\order{\bv{u}}{0}$ would then allow the determination of the elastic stresses induced by the deformation and the consequent feedback on the motion of the surrounding fluid.

\section{Summary}
\label{sec: Discussion}
The \ac{ALE} formulation reported in this article is motivated by a series of articles reported by~\citet{Bradley1996Acoustic-Stream0,bradley2012acoustic}, concerning the differences between the Lagrangian and Eulerian velocity fields inside acoustofluidic devices. As reported by~\citet{bradley2012acoustic}, the Eulerian velocity in these devices exhibit various contradictory features such as the mass flux across an oscillating wall and a slipping flow at the surface of the oscillating wall. These non-physical flow features are not observed in the corresponding Lagrangian flow field. This observation, combined with the fact that the microacoustofluidic flows are usually visualized via the motion of tracer beads that are indicative of the Lagrangian flow field, favors the development of a more readily interpretable formulation based on the Lagrangian mean velocity field.

The formulation reported in this article employs the same perturbation approach reported for Eulerian formulations by several authors \citep[see, e.g.,][]{koster2007numerical,muller2012numerical,nama2015numerical,Vanneste2011Streaming-by-Le0}. However, the proposed \ac{ALE} formulation differs significantly in two aspects: (\emph{i}) the first-order system is formulated in terms of the first-order fluid displacement, thereby offering a natural extension to the more complex fluid-structure interaction problems involving inclusions inside the microchannels; and (\emph{ii}) the primary unknown of our second-order problem is the Lagrangian mean velocity, thereby circumventing the need to employ the calculation of the Stokes drift for the determination of mean particle trajectories as is the case, for example, in experimental validation via particle velocimetry.  Moreover, the proposed \ac{ALE} formulation obtains a second-order problem that is intrinsically steady and has unambiguous boundary conditions.  As we have shown in the result section, our formulation does not exhibit any non-physical features pertaining to Eulerian formulations as discussed by~\citet{Bradley1996Acoustic-Stream0,bradley2012acoustic}. 

We point out that our formulation presupposes the existence of a separation of time scales governed by a parameter $\beta$ whose value is chosen to recover some basic results from traditional Eulerian approaches.  However, our choice of the value of $\beta$ is not the only one that is meaningful as it has been discussed in \citet{Xie2014multiscale}.  Other possible schemes will be examined in future work, which will also include an investigation the influence of slip boundary condition along the lines of~\citet{xie2014boundary}, and an analysis of the \ac{FSI} problem involving a soft deformable solid particle immersed in a streaming flow. 

\section*{Appendix}
\label{Appendix: A}

\subsection{Kinematic Representation Formulas}
In the paper we have used some formulas that are not standard and require some background explanation.  We begin with the derivation of the representation formula for $J_{\bv{\xi}}$, the Jacobian determinant of the tensor $\tensor{F}_{\bv{\xi}}$.  We recall that $\tensor{F}_{\bv{\xi}} = \tensor{I} + \tensor{h}$, where, for typographical convenience we have used $\tensor{h} = \nabla_{\bv{x}} \bv{\xi}$.  With this in mind, we recall that the characteristic polynomial of a generic second-order tensor $\tensor{A}$ has the following representation (see, e.g., \citet{GurtinFried_2010_The-Mechanics_0}):
\begin{equation}
\label{eq: characteristic polynomial}
\det(\tensor{A} - \lambda \tensor{I}) =
- \lambda^{3}
+ \mathscr{I}_{1}(\tensor{A}) \lambda^{2}
- \mathscr{I}_{2}(\tensor{A}) \lambda
+ \mathscr{I}_{3}(\tensor{A})
\end{equation}
where $\mathscr{I}_{1}(\tensor{A})$, $\mathscr{I}_{2}(\tensor{A})$, $\mathscr{I}_{3}(\tensor{A})$, are the principal invariants of $\tensor{A}$ defined as follows:
\begin{equation}
\label{eq: invariants defs}
\mathscr{I}_{1}(\tensor{A}) \coloneqq \trace(\tensor{A}),
\quad
\mathscr{I}_{2}(\tensor{A})
\coloneqq
\tfrac{1}{2}
\bigl[\mathscr{I}_{1}(\tensor{A})^{2}
-
\trace(\tensor{A}^2)
\bigr],
\quad
\mathscr{I}_{3}(\tensor{A}) \coloneqq \det\tensor{A}.
\end{equation}
Observing that the principal invariants, when expressed in terms of the components of the tensor $\tensor{A}$, are polynomials of said components of order $1$, $2$, and $3$, respectively, we have that
\begin{equation}
\mathscr{I}_{1}(\tensor{A}) = \mathcal{O}(\|\tensor{A}\|), \quad
\mathscr{I}_{2}(\tensor{A}) = \mathcal{O}\bigl(\|\tensor{A}\|^{2}\bigr),
\quad\text{and}\quad
\mathscr{I}_{3}(\tensor{A}) = \mathcal{O}\bigl(\|\tensor{A}\|^{3}\bigr).
\end{equation}
Going back to $J_{\bv{\xi}} = \det(\tensor{I} + \tensor{h})$, we then have that
\begin{equation}
\label{eq: representation of Jxi general}
J_{\bv{\xi}} = \det(\tensor{h} - \lambda \tensor{I})|_{\lambda = -1}
\quad\Rightarrow\quad
J_{\bv{\xi}} = 1 + \mathscr{I}_{1}(\tensor{h}) + \mathscr{I}_{2}(\tensor{h}) + \mathscr{I}_{3}(\tensor{h}).
\end{equation}
Next, we turn to the representation formula of the tensor $\tensor{F}_{\bv{\xi}}^{-1}$. We recall that the Cayley-Hamilton theorem, when applied to a second-order tensor, yields
\begin{equation}
\label{eq: Cayley Hamilton}
\tensor{F}_{\bv{\xi}}^{3}
- \mathscr{I}_{1}(\tensor{F}_{\bv{\xi}}) \tensor{F}_{\bv{\xi}}^{2}
+ \mathscr{I}_{2}(\tensor{F}_{\bv{\xi}}) \tensor{F}_{\bv{\xi}}
- \mathscr{I}_{3}(\tensor{F}_{\bv{\xi}}) \tensor{I} = \tensor{0}.
\end{equation}
As $\tensor{F}_{\bv{\xi}}$ is invertible, we then have
\begin{equation}
\label{eq: Cayley Hamilton solution}
\mathscr{I}_{3}(\tensor{F}_{\bv{\xi}})\tensor{F}_{\bv{\xi}}^{-1}
=
\tensor{F}_{\bv{\xi}}^{2}
- \mathscr{I}_{1}(\tensor{F}_{\bv{\xi}}) \tensor{F}_{\bv{\xi}}
+ \mathscr{I}_{2}(\tensor{F}_{\bv{\xi}}) \tensor{I}.
\end{equation}
Recalling that $\tensor{F}_{\bv{\xi}} = \tensor{I} + \tensor{h}$ and using the definitions for the principal invariants of a tensor, one can expand and simplify the right-hand side of Eq.~\eqref{eq: Cayley Hamilton solution}, so that this equation can be rewritten as
\begin{equation}
\label{eq: Cayley Hamilton solution simplified}
(
1
+
\mathscr{I}_{1}(\nabla_{\bv{x}}\bv{\xi})
+
\mathscr{I}_{2}(\nabla_{\bv{x}}\bv{\xi})
+
\mathscr{I}_{3}(\nabla_{\bv{x}}\bv{\xi})
)
\tensor{F}_{\bv{\xi}}^{-1}
=
\tensor{I}
-
\tensor{h} + \mathscr{I}_{1}(\tensor{h}) \tensor{I}
+
\tensor{h}^{2} - \mathscr{I}_{1}(\tensor{h}) \tensor{h} + \mathscr{I}_{2}(\tensor{h}) \tensor{I}.
\end{equation}
Now, we consider an additive decomposition of $\tensor{F}_{\bv{\xi}}^{-1}$ in terms of contributions of various orders in $\|\tensor{h}\|$, i.e., $\tensor{F}_{\bv{\xi}}^{-1} = \order[-1]{\tensor{F}}{0}_{\bv{\xi}} + \order[-1]{\tensor{F}}{1}_{\bv{\xi}} + \cdots$.  As we are interested only in the contributions to $\tensor{F}_{\bv{\xi}}^{-1}$ of order $0$, $1$, and $2$, setting equal the orders $0$, $1$, and $2$ of the left-hand side of Eq.~\eqref{eq: Cayley Hamilton solution simplified} with the corresponding orders of the right-hand side of the same equation, we then obtain the following result:
\begin{equation}
\label{eq: fxi inverse orders 0 and 1}
\order[-1]{\tensor{F}}{0}_{\bv{\xi}} = \tensor{I},
\quad
\order[-1]{\tensor{F}}{1}_{\bv{\xi}} = -\tensor{h},
\quad\text{and}\quad
\order[-1]{\tensor{F}}{2}_{\bv{\xi}} = \tensor{h}^{2}.
\end{equation}

\subsection{On the velocity decomposition into first- and second-order contributions}
Referring to Eq.~\eqref{eq: *notation}, for the full velocity field we have
\begin{equation}
\label{eq: app vL classic 1}
\bv{v}(\bv{y},t)\Bigm \lvert_{\bv{y}=\bv{x}+\xi(\bv{x},t)}=\bv{v}^{*}(\bv{x},t)
\end{equation}
Expanding the LHS of this equation to the second-order accuracy in a neighborhood of $\bv{y} = \bv{x}$, we have:
\begin{equation} 
\label{eq: app vL classic 2}
\bv{v}(\bv{x},t)+\nabla_{\bv{x}}\bv{v}^{*}\tensor{F}_{\bv{\xi}}^{-1}\bv{\xi}(\bv{x},t)=\bv{v}^{*}(\bv{x},t).
\end{equation}
In the non-ALE approach, it is $\bv{v}(\bv{x},t)$ that is split into the first- and second-order components, while in the ALE approach it is the RHS that is subject to such decomposition. This gives:
\begin{equation}
\label{eq: app vL classic 3}
\order{\bv{v}}{1}(\bv{x},t) + \order{\bv{v}}{2}(\bv{x},t) 
+
\nabla_{\bv{x}}\bigl(\order[*]{\bv{v}}{1} + \order[*]{\bv{v}}{2} \bigr)\tensor{F}_{\bv{\xi}}^{-1} \bv{\xi}(\bv{x},t) = \order[*]{\bv{v}}{1}(\bv{x},t) + \order[*]{\bv{v}}{2}(\bv{x},t).
\end{equation}
Matching orders, we have that the first-order velocities for the traditional Eulerian formulation and our \ac{FSI} formulation coincide:
\begin{equation}
\label{eq: app vL classic 4}
\order{\bv{v}}{1}(\bv{x},t) = \order[*]{\bv{v}}{1}(\bv{x},t).
\end{equation}
However, for the the second-order velocities we have
\begin{equation}
\label{eq: app vL classic 5}
\order{\bv{v}}{2} + \nabla_{\bv{x}}\order{\bv{v}}{1}\tensor{F}_{\bv{\xi}}^{-1} \bv{\xi} = \order[*]{\bv{v}}{2},
\end{equation}
where, for convenience, we have omitted the argument `$(\bv{x},t)$'.  To the second-order, and upon time averaging, this amounts to
\begin{equation}
\label{eq: velocityrelation}
\langle\order{\bv{v}}{2}\rangle + \langle \nabla_{\bv{x}}\order{\bv{v}}{1} \bv{\xi}\rangle = \LM{\bv{v}},
\end{equation}
where the second term on the left-hand side of the above expression is the Stokes drift.

Similar comparisons can be derived for other quantities of interest, such as the density or the stress.


\section*{Acknowledgment}
This work was supported by National Institutes of Health (1R01 GM112048-01A1, 1R33EB019785-01), National Science Foundation (CBET-1438126 and IIP-1346440), and the Penn State Center for Nanoscale Science (MRSEC) under grant DMR-0820404.

\bibliography{NamaEtAlAcoustoFluidicsFSI}
\bibliographystyle{jfm}

\end{document}